\documentclass[12pt]{article}

\usepackage{amsmath}
\usepackage{amssymb}
\usepackage[pdftex]{graphicx}
\usepackage{amscd}
\usepackage{pictexwd,dcpic}
\usepackage{slashed}
\usepackage{mathtools}
\usepackage{braket}

\numberwithin{equation}{section}

\begin{document}

\begin{flushright}
UTTG-06-16
\end{flushright}

\begin{center}
\LARGE{Topological String Theory Revisited I: The Stage}

\large{Bei Jia}

\footnotesize{Theory Group, University of Texas at Austin, Austin, TX 78712, USA}

\footnotesize{beijia@physics.utexas.edu}

\end{center}

\date{}


\begin{abstract}
In this note we reformulate topological string theory using supermanifolds and supermoduli spaces, following the approach worked out by Witten for superstring perturbation theory in arXiv:1209.5461. We intend to make the construction geometrical in nature, by using supergeometry techniques extensively. The goal is to establish the foundation of studying topological string amplitudes in terms of integration over appropriate supermoduli spaces.
\end{abstract}

\newpage

\tableofcontents

\newpage

\section{Introduction}

Superstring perturbation theory is most naturally defined using techniques from the theory of supermanifolds. Recently, there has been many works on an elegant and precise formulation of superstring perturbation theory in terms of integrations over supermoduli spaces of $\mathcal{N}=1$ super Riemann surfaces \cite{witten-spt}. In particular, it was shown that most of such supermoduli spaces are not projected \cite{donagi-witten-proj, donagi-witten-superatiyah}, meaning that we should not hope to naively ``integrate out odd moduli'' and formulate superstring amplitudes as integrations over ordinary moduli spaces of Riemann surfaces. As such, these supermoduli spaces are more than just glorified vector bundles over bosonic moduli spaces and worth studying by themselves.

There is a close relative to superstring theory, namely topological string theory. There are many existing approaches (see, for example, \cite{witten-ts, dvv, distler-nelson, dislter-nelson-2, bcov}) towards defining such a theory. Morally speaking, it is a combination of topological field theory and worldsheet topological gravity, in an appropriate sense. In some approaches, the topological properties of topological string theory are given by definition \cite{witten-ts}. There are also geometrical formulations using various supermanifold techniques \cite{dvv, distler-nelson, dislter-nelson-2} that precede what we will discuss in this paper. There are also approaches that build topological string theory by modeling bosonic string theory, utilizing the properties of the topological field theory that lives on worldsheet \cite{bcov}.

In this paper, we will be following the approach pioneered in \cite{distler-nelson, dislter-nelson-2}. In those papers, it was found that topological string worldsheet can be most naturally constructed as semirigid super Riemann surfaces (SRSS). These are supermanifolds of complex dimension $1|2$, i.e. one bosonic dimension and two fermionic dimension. They can be obtained from performing topological twists on $\mathcal{N}=2$ super Riemann surfaces in a geometrical sense. Then it is only natural to define topological string amplitudes as integrations over the supermoduli spaces of such semirigid super Riemann surfaces.

In sight of this approach, it is natural to ask whether this is necessary. We will argue that, analogous to the reasoning in \cite{witten-spt}, that we should define things on supermoduli spaces if these spaces are not projected\footnote{The proof of the statement that these supermoduli spaces are not projected will appear in a follow-up paper \cite{distler-jia-2}.}. So the situation is quite similar to superstring perturbation theory. We should not hope to obtain sensible results in general because it is not valid to integrate out odd moduli generically.

The aim of the current notes is to lay out the foundation of such an approach, using techniques from supergeometry. In section 2, we will first review the construction and properties of $\mathcal{N}=2$ super Riemann surfaces. Through doing this, we will introduced many concepts, such as superconformal vector fields and superconformal coordinate transformations, which are necessary for later discussions. In section 3 we move on to define semirigid super Riemann surfaces, by twisting $\mathcal{N}=2$ super Riemann surfaces. Along the way we formulate topological A and B model field theories on such semirigid super Riemann surfaces. We also discuss some properties of associated supermoduli spaces, which are central players of our story. Finally in section 4 we build our main objects, namely topological string amplitudes. We formulate everything in terms of supergeometry, and define topological string amplitudes as integrations over supermoduli spaces of semirigid super Riemann surfaces. 

There are several appendices. In appendix \ref{sm} we review some basics about supermanifolds in general. In appendix \ref{berezinian} we study and prove some properties of Berezinian line bundles of various supermanifolds that appeared in the main text. In appendix \ref{02} we construct worldsheets with $(0,2)$ supersymmetry. In appendix \ref{deformation} we study general deformation theory of $\mathcal{N}=2$ super Riemann surfaces. In appendix \ref{alternative} we give an alternative way of defining topological string amplitudes using integral forms.

\section{$\mathcal{N} = 2$ Worldsheet}

\subsection{Review of $\mathcal{N} = 2$ Super Riemann Surfaces}  \label{srs}
\subsubsection{The Stage}
An $\mathcal{N} = 2$ super Riemann surface $S$ \cite{witten-srs, cohn-n=2} is a complex supermanifold of dimension $1|2$, with the so called $\mathcal{N} = 2$ superconformal structure. We denote the local coordinates of $S$ as $(z|\theta^-,\theta^+)$. Let $TS$ be the holomorphic tangent bundle of $S$. A $\mathcal{N} = 2$ superconformal structure requires that there exists two odd subbundles, denoted as $\mathcal{D}_-$ and $\mathcal{D}_+$, of $TS$ with rank $0|1$. It is further required that $\mathcal{D}_-$ and $\mathcal{D}_+$ to be integrable. Here, the integrability means that, for any given section $D_-$ of  $\mathcal{D}_-$, we have $D_-^2 = f(z|\theta^-,\theta^+) D_-$ for some function $f(z|\theta^-,\theta^+) $. Similarly,  for any given section $D_+$ of  $\mathcal{D}_+$, we have $D_+^2 = g(z|\theta^-,\theta^+) D_+$ for some function $g(z|\theta^-,\theta^+) $. Finally, we need the sections of $\mathcal{D}_- \otimes \mathcal{D}_+$ to be everywhere linearly independent of sections of $\mathcal{D}_-$ and $\mathcal{D}_+$.

One can prove that there exist a local coordinates system $(z|\theta^-,\theta^+)$ on $S$, called superconformal coordinates, in which the generators of the sections of $\mathcal{D}_-$ and $\mathcal{D}_+$ can be expressed as
\begin{equation} \label{superderivative}
\begin{split}
D_- & = \frac{\partial}{\partial \theta^-} + \theta^+ \frac{\partial}{\partial z}, \\
D_+ & = \frac{\partial}{\partial \theta^+} + \theta^- \frac{\partial}{\partial z}.
\end{split}
\end{equation}
It is then straightforward to check that
\begin{equation}
D_-^2 = D_+^2 = 0, \ \ \ \{D_-, D_+\} = 2\partial_z.
\end{equation}
These are precisely the usual super derivatives that are commonly used for (0,2) supersymmetry on $\mathbb{R}^2$.

Let $\Phi: S \rightarrow \mathbb{C}$ be a complex function on $S$. If $\Phi$ is annihilated by $D_-$, then it is called a chiral function. In turn, if another complex function $\widetilde{\Phi}$ is annihilated by by $D_+$, then it is called an antichiral function\footnote{If we study (0,2) supersymmetric theories (which we study in appendix \ref{02}), then $\Phi$ and $\widetilde{\Phi}$ are precisely the so called (0,2) chiral and antichiral superfields.}. 

One can show \cite{witten-srs} that there are natural projections
\begin{equation*}
\begindc{\commdiag}[50]
\obj(1,1)[s]{$S$}
\obj(0,0)[x]{$X$}
\obj(2,0)[x']{$X'$}
\mor{s}{x}{$\pi$}
\mor{s}{x'}{$\pi'$}
\enddc
\end{equation*}
where $X$ and $X'$ are complex supermanifold of dimension $1|1$ without any additional structure, with local coordinates $(z - \theta^-\theta^+|\theta^+)$ and $(z - \theta^+\theta^-|\theta^-)$, respectively. Then a chiral function is $\Phi = \pi^* \varphi$, the pullback of a complex function $\varphi: X \rightarrow \mathbb{C}$. Similarly, an antichiral function is $\widetilde{\Phi} = \pi'^* \varphi'$, the pullback of a complex function $\varphi': X' \rightarrow \mathbb{C}$.

To see this, notice that equation (\ref{superderivative}) can be rewritten as
\begin{equation}
\begin{split}
D_- & = e^{-\theta^-\theta^+\partial_z} \frac{\partial}{\partial \theta^-} e^{\theta^-\theta^+\partial_z},\\
D_+ & = e^{\theta^-\theta^+\partial_z}\frac{\partial}{\partial \theta^+} e^{-\theta^-\theta^+\partial_z}.
\end{split}
\end{equation}
Therefore, functions annihilated by $D_-$ and $D_+$ are of the form
\begin{equation} 
\begin{split}
e^{-\theta^-\theta^+\partial_z}\Phi(z|\theta^+) & =  \Phi(z-\theta^-\theta^+|\theta^+), \\
e^{\theta^-\theta^+\partial_z}\widetilde{\Phi}(z|\theta^-) & =  \widetilde{\Phi}(z-\theta^+\theta^-|\theta^-). 
\end{split}
\end{equation}
so are pullbacks of functions on $X$ and $X'$, respectively.

Geometrically, $X$ is the space of orbits generated by sections of any nonzero section of $\mathcal{D}_-$, while $X'$ is the space of orbits generated by sections of $\mathcal{D}_+$. In other words:
\begin{equation}
X = S / \mathcal{D}_-, \ \ \ \ X' = S / \mathcal{D}_+.
\end{equation}
One can show that the holomorphic vector fields on a generic complex supermanifold of dimension $1|1$ (without any additional structure) in fact generate $\mathcal{N}=2$ superconformal algebra, which naturally acts on chiral and antichiral functions. That is to say, the general coordinate transformations on $X$ (or $X'$) contain the exact same data as the superconformal coordinate transformations on $S$. Therefore, the supermoduli space of a $\mathcal{N} = 2$ super Riemann surface is isomorphic to the supermoduli space of a $1|1$ dimensional complex supermanifold without any additional structure.

Note that we do not ask $\widetilde{\Phi}$ to be the complex conjugate of $\Phi$. In particular, we are not allowed to take $\theta^+$ to be the complex conjugate of $\theta^-$. This is because, in Euclidean signature, there is no reality condition on spinors. In contrast, in Lorentzian signature (when it exists), we can in general put real structures on spinors as usually done in the literature.

In terms of local superconformal coordinates on a coordinate patch, we have the usual expansion
\begin{equation}
\begin{split}
\Phi & = \phi + \theta^+ \psi - \theta^- \theta^+ \partial_z \phi, \\
\widetilde{\Phi} & = \widetilde{\phi} + \theta^- \psi' + \theta^- \theta^+ \partial_z \widetilde{\phi}.
\end{split}
\end{equation}
where $\phi$ and $\widetilde{\phi}$ are scalar fields, while $\psi$ and $\psi'$ are Weyl fermionic fields.

\subsubsection{Superconformal Transformations}

Among all the holomorphic vector fields on $S$, there is a special class called superconformal vector fields, which are defined to preserve odd line bundles  $\mathcal{D}_-$ and $\mathcal{D}_+$ \cite{witten-srs}. They generate $\mathcal{N}=2$ superconformal algebra. In terms of local superconformal coordinates $(z|\theta^-,\theta^+)$, a basis of superconformal vetor fields can be written as
\begin{equation} \label{superconformal-vector}
\begin{split}
\mathcal{T} & =  g(z)  \partial_z + \frac{1}{2} \partial_z g(z)  (\theta^-\partial_{\theta^-} + \theta^+\partial_{\theta^+}), \\
\mathcal{J} & = k(z)  (\theta^-\partial_{\theta^-} - \theta^+\partial_{\theta^+}), \\
\mathcal{G}^- & =\left (\alpha^-(z) + \theta^-\theta^+ \partial_z \alpha^-(z) \right) \partial_{\theta^-} -  \alpha^-(z) \theta^+ \partial_z, \\
\mathcal{G}^+ & = \left( \alpha^+(z) -  \theta^-\theta^+ \partial_z \alpha^+(z) \right) \partial_{\theta^+} -  \alpha^+(z) \theta^- \partial_z,
\end{split}
\end{equation}
where $g(z)$ and $k(z)$ are even functions of $z$, while $\alpha^-(z)$ and $\alpha^+(z)$ are odd functions of $z$. 

Notice that $g(z)$, $k(z)$, $\alpha^-(z)$ and $\alpha^+(z)$ are all functions of $z$ only. We in fact can combine them into a superfield
\begin{equation}  \label{superconformal-field}
\mathcal{Y}(z|\theta^{\mp}) = g + 2 \theta^- \alpha^+ +   2 \theta^+ \alpha^- - 2 \theta^- \theta^+ k.
\end{equation}
We call $\mathcal{Y}(z|\theta^{\mp})$ a superconformal superfield. Then a general superconformal vector field can be put into the following form:
\begin{equation}
\mathcal{W} = \mathcal{Y}(z|\theta^{\mp}) \partial_z + \frac{1}{2} D_- \mathcal{Y} D_+ + \frac{1}{2} D_+ \mathcal{Y} D_-.
\end{equation}
It is then straightforward to check that
\begin{equation}
\mathcal{W} = \mathcal{T} + \mathcal{J} + \mathcal{G}^- + \mathcal{G}^+.
\end{equation}
Therefore, we see that there is a one-to-one map between superconformal vector fields and superfields like $\mathcal{V}(z|\theta^{\mp})$.

We should emphasize that $\mathcal{T}, \mathcal{J}, \mathcal{G}^{\mp}$ above are vector fields on a $\mathcal{N} = 2$ super Riemann surface $S$. As such, they are the symmetry generators in the geometric sense. As we shall see in a moment, these superconformal vector fields do indeed generate the usual $\mathcal{N} = 2$ superconformal algebra with zero central charge. This is natural since they generate the superconformal coordinate transformations of $S$, which is defined to have $\mathcal{N} = 2$ superconformal structure. They should not be confused with the usual field theoretical generators of $\mathcal{N} = 2$ superconformal algebra in conformal field theories. This is because, in a general $\mathcal{N} = 2$ superconformal field theory, the corresponding superconformal symmetry generators (usually denoted as $T, J, G^{\mp}$) are functionals of the field operators in this field theory. In other words, as usually happens in field theory, we obtain a representation of $\mathcal{N} = 2$ superconformal algebra on the Hilbert space of the given superconformal field theory (a homomorphism between the $\mathcal{N} = 2$ superconformal algebra generated by $\mathcal{T}, \mathcal{J}, \mathcal{G}^{\mp}$ and the operator algebra on this Hilbert space).

One should be able to check explicitly that the commutators/anticommutators between these superconfomal vector fields and $D_-$, $D_+$ are proportional to $D_-$, $D_+$. It is indeed the case:
\begin{equation} \label{superconformal-vector-act-D}
\begin{split}
[\mathcal{T}, D_{-}] & = -\frac{1}{2} \left( \partial_z g - \theta^-\theta^+ \partial_z\partial_z g\right) D_{-}, \\
[\mathcal{T}, D_{+}] & = -\frac{1}{2} \left( \partial_z g + \theta^-\theta^+ \partial_z\partial_z g\right) D_{+}, \\
[\mathcal{J}, D_{-}] & = - \left(k - \theta^-\theta^+ \partial_z k \right) D_{-}, \\
[\mathcal{J}, D_{+}] & = \left(k + \theta^-\theta^+ \partial_z k \right) D_{+}, \\
\{ \mathcal{G}^-, D_- \} & = 2 \theta^+ \partial_z \alpha^- D_{-}, \\
\{ \mathcal{G}^+, D_+ \} & = 2 \theta^- \partial_z \alpha^+ D_{+}, \\
\{ \mathcal{G}^-, D_+ \} & = \{ \mathcal{G}^+, D_- \} = 0.
\end{split}
\end{equation}
In other words, this explicitly shows that our $\mathcal{N}=2$ superconformal vector fields $\mathcal{T}, \mathcal{J}, \mathcal{G}^{\mp}$ indeed preserve the odd line bundles $\mathcal{D}_-$ and $\mathcal{D}_+$. 

Here are the expansion modes of the above basis of superconformal vector fields (with possible poles at $z=0$):
\begin{equation} \label{superconformal-generator}
\begin{split}
\mathcal{L}_n & = - z^{n+1}  \partial_z - \frac{1}{2}  (n+1) z^{n} (\theta^-\partial_{\theta^-} + \theta^+\partial_{\theta^+}), \\
\mathcal{J}_n & = z^{n}  (\theta^-\partial_{\theta^-} - \theta^+\partial_{\theta^+}), \\
\mathcal{G}^-_r & =\left ( z^{r+1/2} + \theta^-\theta^+ \left(r+\frac{1}{2}\right) z^{r-1/2} \right) \partial_{\theta^-} - z^{r+1/2} \theta^+ \partial_z, \\
\mathcal{G}^+_r & = \left( z^{r+1/2} -  \theta^-\theta^+ \left(r+\frac{1}{2}\right) z^{r-1/2} \right) \partial_{\theta^+} - z^{r+1/2} \theta^- \partial_z,
\end{split}
\end{equation}
where $n \in \mathbb{Z}$, while $r \in \mathbb{Z} + \frac{1}{2}$. Note there are no poles when $n\ge 0$ and $r\ge 1/2$. A simple computation shows that these generators produce the usual $\mathcal{N} = 2$ superconformal algebra in the Neveu-Schwarz (NS) sector with zero central charge:
\begin{equation} \label{n=2-algebra}
\begin{split}
[\mathcal{L}_m, \mathcal{L}_n] & =  (m - n) \mathcal{L}_{m+n}, \\
[\mathcal{L}_m, \mathcal{G}^{\pm}_r] & = \left(\frac{m}{2} - r \right) \mathcal{G}^{\pm}_{m+r}, \\
[\mathcal{L}_m, \mathcal{J}_n] & = - n \mathcal{J}_{m+n}, \\
\{ \mathcal{G}^-_r, \mathcal{G}^+_s \} & = 2 \mathcal{L}_{r+s} + (r-s) \mathcal{J}_{r+s},\\
\{ \mathcal{G}^-_r, \mathcal{G}^-_s \} & = \{ \mathcal{G}^+_r, \mathcal{G}^+_s \}  = [\mathcal{J}_m, \mathcal{J}_n]  = 0, \\
[\mathcal{J}_m, \mathcal{G}^-_r] & = \mathcal{G}^-_{r+m}, \\
[\mathcal{J}_m, \mathcal{G}^+_r] & = - \mathcal{G}^+_{r+m}. \\ 
\end{split}
\end{equation}
The absence of a nonzero central charge here is exactly what one should expect. In order to have conformal symmetry at the full quantum level, the total central charge has to vanish.

Let's remark that $\mathcal{G}^{\mp}$ anticommute with $D_{\mp}$ when $\alpha^{\mp}$ are constants. This is what one would expect because in this case $\mathcal{G}^{\mp}$ are proportional to the usual rigid supersymmetry generators $\mathcal{Q}_{\mp}$ in any $\mathcal{N}=2$ supersymmetric (not necessarily conformal) quantum field theories:
\begin{equation} \label{supercharge}
\begin{split}
\mathcal{Q}_- & = \frac{\partial}{\partial \theta^-} - \theta^+ \frac{\partial}{\partial z}, \\
\mathcal{Q}_+ & = \frac{\partial}{\partial \theta^+} - \theta^- \frac{\partial}{\partial z}. 
\end{split}
\end{equation}
In particular, from the expansion modes in (\ref{superconformal-generator}) we clearly see that 
\begin{equation}
\mathcal{Q}_-  = \mathcal{G}^-_{-1/2}, \ \ \mathcal{Q}_+  = \mathcal{G}^+_{-1/2}.
\end{equation}
This is a reflection on the fact that superconformal symmetry contains rigid supersymmetry.

Now let's discuss the superconformal coordinate transformations generated by these superconformal vector fields. Let $(z|\theta^-,\theta^+)$ and $(z'|{\theta^-}',{\theta^+}')$ be two sets of local superconformal coordinates. Then the superconformal coordinate transformations between them are given by \cite{cohn-n=2, distler-nelson}
\begin{equation} \label{superconformal-transf}
\begin{split}
z' & = f(z) + \theta^- a(z) \rho(z) + \theta^+ e(z) \alpha(z) + \theta^- \theta^+ \partial_{z} (\alpha(z) \rho(z)), \\
{\theta^-}^{\prime} & = \alpha(z) + \theta^- a(z) + \theta^- \theta^+ \partial_z \alpha(z), \\
{\theta^+}^{\prime} & = \rho(z) + \theta^+ e(z) - \theta^- \theta^+ \partial_z \rho(z) ,
\end{split}
\end{equation}
with the following constraint
\begin{equation}
e a = \partial_z f + \alpha \partial_z \rho + \rho \partial_z \alpha,
\end{equation}
where $f,\ a, \ e$ are even functions of $z$, while $\alpha, \ \rho$ are odd functions of $z$. These superconformal transformations are generated by the above superconformal vector fields $\mathcal{T}, \mathcal{J}, \mathcal{G}^{\mp}$. By counting, we see that we indeed have two independent even generators as well as two independent odd generators. One can think of a $\mathcal{N}=2$ super Riemann surface as constructed by gluing together pieces of $\mathbb{C}^{1|2}$ via these superconformal coordinate transformations.

The corresponding local sections $D_-$, $D_+$ and $D_-'$, $D_+'$ are related to each other by
\begin{equation}
D_- = ( D_-{\theta^-}') D_-', \ \ D_+ = ( D_+{\theta^+}') D_+'.
\end{equation}
which means the odd line bundles $\mathcal{D}_-$ and $\mathcal{D}_+$ are preserved by these superconformal coordinate transformations, as they should be.

\subsection{(2,2) Worldsheet}  \label{22}

One peculiar property of two-dimensional theories is that chiral fermions in two dimensions fall into two independent types: left-movers and right-movers. Using complex coordinates, they correspond to fermions that are holomorphic and antiholomorphic respectively. In \cite{witten-srs, witten-spt}, Witten introduced an explicit way of characterizing these two different degrees of freedom. For example, the worldsheet of a heterotic string is given by
\begin{equation}
\Sigma_{(0,1)} \hookrightarrow \Sigma_L \times \Sigma_R,
\end{equation}
where $\Sigma_L$ is an ordinary Riemann surface, while $\Sigma_R$ is an $\mathcal{N}=1$ super Riemann surface. Then $\Sigma_{(0,1)}$ is a smooth submanifold\footnote{In this paper, all smooth supermanifolds are cs manifolds, where cs stands for complex supersymmetric \cite{cs}. In a nutshell, this means that a cs manifold has a reduced space which is a real smooth manifold; we can't ask for any reality conditions once we include the odd coordinates.} of dimension $2|1$ which is close to the ``diagonal'', meaning that ${\Sigma_{(0,1)}}_{\text{red}}$ is sufficiently close to the diagonal of $(\Sigma_L \times \Sigma_R)_{\text{red}}$. We can interpret $\Sigma_L \times \Sigma_R$ as the complexification of $\Sigma_{(0,1)}$.

Now we apply the same trick again to formulate (2,2) supersymmetric theories. Let's consider 
\begin{equation}
\Sigma_{(2,2)} \hookrightarrow S_L \times S_R,
\end{equation}
where both $S_L$ and $S_R$ are $\mathcal{N}=2$ super Riemann surfaces, and $\Sigma_{(2,2)}$ is a smooth submanifold of dimension $2|4$.

Let's parametrize $S_L$ with local superconformal coordinate $(\widetilde{z}|\widetilde{\theta}^-,\widetilde{\theta}^+)$, and $S_R$ with local superconformal coordinates $(z|\theta^-,\theta^+)$. Then the holomorphic structure on $S_L$ gives rise to the antiholomorphic structure on $\Sigma_{(2,2)}$, while the holomorphic structure on $S_R$ gives rise the the holomorphic structure on $\Sigma_{(2,2)}$. In other words, we parametrize $\Sigma_{(2,2)}$ by $(\widetilde{z}, z|\widetilde{\theta}^-,\widetilde{\theta}^+, \theta^-,\theta^+)$.

As shown in appendix \ref{berezinian}, the Berezinian line bundle of a $\mathcal{N}=2$ super Riemann surface is canonically trivial. Then the isomorphism
\begin{equation}
\text{Ber}(\Sigma_{(2,2)}) \cong Ber(S_L \times S_R)|_{\Sigma_{(2,2)}}
\end{equation}
shows that the Berezinia line bundle of $\Sigma_{(2,2)}$ (as a smooth supermanifold) is also trivial. Therefore, any function on $\Sigma_{(2,2)}$ can be integrated over $\Sigma_{(2,2)}$.

For our purposes, we would like to study a special class of functions on $\Sigma_{(2,2)}$, typically called (2,2) chiral superfields. As we have discussed, there are superconformal coordinates $(\widetilde{z}|\widetilde{\theta}^-,\widetilde{\theta}^+)$ on $S_L$, and superconformal coordinates $(z|\theta^-,\theta^+)$ on $S_R$, such that
\begin{equation} \label{22superderivative}
\begin{split}
D_- & = \frac{\partial}{\partial \theta^-} + \theta^+ \frac{\partial}{\partial z}, \\
D_+ & = \frac{\partial}{\partial \theta^+} + \theta^- \frac{\partial}{\partial z}, \\
\widetilde{D}_- & = \frac{\partial}{\partial \widetilde{\theta}^-} +  \widetilde{\theta}^+ \frac{\partial}{\partial \widetilde{z}}, \\
\widetilde{D}_+ & = \frac{\partial}{\partial \widetilde{\theta}^+} +  \widetilde{\theta}^- \frac{\partial}{\partial \widetilde{z}},
\end{split}
\end{equation}
which generate sections of odd line bundles $\widetilde{\mathcal{D}}_-, \ \widetilde{\mathcal{D}}_+ \rightarrow S_L$ and $\mathcal{D}_-, \ \mathcal{D}_+ \rightarrow S_R$, respectively. We can then restrict these bundles to the ``diagonal'' $\Sigma_{(2,2)}$. Then a (2,2) chiral superfield\footnote{The word ``chiral'' here is borrowed from four-dimensional $\mathcal{N}=1$ supersymmetry; it doesn't mean two-dimensional chirality, in contrast to (0,2) chiral superfields.} is a complex function $\Phi: \Sigma_{(2,2)} \rightarrow \mathbb{C}$ satisfying
\begin{equation}
D_- \Phi = \widetilde{D}_- \Phi = 0.
\end{equation}
Similarly, an antichiral superfield is a complex function $\widetilde{\Phi}: \Sigma_{(2,2)} \rightarrow \mathbb{C}$ satisfying
\begin{equation}
D_+ \widetilde{\Phi} = \widetilde{D}_+ \widetilde{\Phi} = 0.
\end{equation}

In terms of local coordinates on a coordinate patch, we have the usual expansion
\begin{equation}
\begin{split}
\Phi  =& \ \phi + \theta^+ \psi + \widetilde{\theta}^+ \lambda + \theta^+ \widetilde{\theta}^+ F - \theta^-\theta^+ \partial_z \phi - \widetilde{\theta}^- \widetilde{\theta}^+ \partial_{\widetilde{z}}\phi \\
& - \theta^+ \widetilde{\theta}^- \widetilde{\theta}^+ \partial_{\widetilde{z}} \psi - \theta^- \theta^+ \widetilde{\theta}^+ \partial_z\lambda + \theta^- \theta^+ \widetilde{\theta}^- \widetilde{\theta}^+ \partial_z \partial_{\widetilde{z}} \phi, \\
\widetilde{\Phi} =& \ \widetilde{\phi} + \theta^- \psi' + \widetilde{\theta}^- \lambda' +  \theta^- \widetilde{\theta}^- F'  + \theta^-\theta^+ \partial_{\widetilde{z}} \widetilde{\phi} + \widetilde{\theta}^- \widetilde{\theta}^+ \partial_z\widetilde{\phi} \\
&+ \theta^- \widetilde{\theta}^- \widetilde{\theta}^+ \partial_z \psi' + \theta^- \theta^+ \widetilde{\theta}^- \partial_{\widetilde{z}}\lambda' + \theta^- \theta^+ \widetilde{\theta}^- \widetilde{\theta}^+ \partial_z \partial_{\widetilde{z}} \widetilde{\phi},
\end{split}
\end{equation}
where where $\phi$ and $\widetilde{\phi}$ are scalar fields, $\psi$ and $\psi'$ are right-moving Weyl fermionic fields, $\lambda$ and $\lambda'$ are left-moving Weyl fermionic fields, and $F, \ F'$ are auxillary fields.

Note now we have two sets of $\mathcal{N}=2$ superconfromal vector fields: $\widetilde{\mathcal{T}}, \widetilde{\mathcal{J}}, \widetilde{\mathcal{G}}^{\pm}$ from $S_L$ and $\mathcal{T}, \mathcal{J}, \mathcal{G}^{\pm}$ from $S_R$. In other words, we have $\mathcal{N}=2$ superconfromal symmetry on both left and right moving degrees of freedom.  One can obtain the usual superconformal transformations between component fields by acting $\widetilde{\mathcal{G}}^{\pm}, \mathcal{G}^{\pm}$ on $\Phi$ and $\widetilde{\Phi}$.

Then the kinetic action for chiral superfields is simply
\begin{equation}
\int_{\Sigma_{(2,2)}} \mathcal{D}(\widetilde{z},z|\widetilde{\theta}^-,\widetilde{\theta}^+, \theta^-,\theta^+) K(\widetilde{\Phi}^{\bar{\imath}}, \Phi^i),
\end{equation}
where $K(\widetilde{\Phi}^{\bar{\imath}}, \Phi^i)$ is a real function. Again, in Euclidean signature we are not allowed to ask $\widetilde{\Phi}^{\bar{\imath}}$ to be the complex conjugate of $\Phi^i$. $K(\widetilde{\Phi}^{\bar{\imath}}, \Phi^i)$ corresponds to the K\"{a}hler potential in Lorentzian theories. That said, we do have a reality condition when we set the odd coordinates to zero, i.e. when we restrict to the reduced space $\Sigma$. Then we can demand $\widetilde{\phi}^i$ to be the complex conjugate of $\phi^i$, where $\widetilde{\phi}^{\bar{\imath}}$ and $\phi^i$ are the lowest components of $\widetilde{\Phi}^{\bar{\imath}}$ and $\Phi^i$, respectively. In that sense, we can think of $\phi^i$ and $\widetilde{\phi}^{\bar{\imath}}$ as the holomorphic and antiholomorphic local coordinates on a K\"{a}hler manifold $M$ with K\"{a}hler potential $K(\widetilde{\phi}^{\bar{\imath}}, \phi^i)$. In that sense, this is the usual (2,2) supersymmetric nonlinear sigma model. 

Let's remark that, since we started with a superconformal structure on our worldsheet, we would like to have the full $(2,2)$ superconformal symmetry in out theory. As such, we will need the target space to be Calabi-Yau. This can be seen in two ways. One traditional way is to compute the one-loop beta function of our sigma model and obtain the Ricci flat condition on $M$. The other way is more transparent in our setup: we started with the full $\mathcal{N} = 2$ superconformal algebra, which contains U(1) current $J$. This means that our theory should have anomaly free left and right moving U(1) R-symmetries. This in turn implies the target space $M$ has to be Calabi-Yau.

In this nonlinear sigma model, we can write down explicit expressions for the generators of the $\mathcal{N} = 2$ superconformal algebra $T_{M}, J_{M}, G^{\mp}_M$:
\begin{equation} \label{nlsm-superconformal-vector}
\begin{split}
T_M & =  -g_{i\bar{\jmath}} \partial_z \phi^i \partial_z \phi^{\bar{\jmath}} + \frac{1}{2} g_{i\bar{\jmath}} \psi^i \partial_z \psi^{\bar{\jmath}} + \frac{1}{2} g_{i\bar{\jmath}} \psi^{\bar{\jmath}}  \partial_z \psi^i, \\
J_M & = \frac{1}{4} g_{i\bar{\jmath}} \psi^i \psi^{\bar{\jmath}}, \\
G^-_M & =\frac{1}{2} g_{i\bar{\jmath}} \psi^{\bar{\jmath}}  \partial_z \phi^i, \\
G^+_M & = \frac{1}{2} g_{i\bar{\jmath}} \psi^i \partial_z \phi^{\bar{\jmath}}.
\end{split}
\end{equation}
The algebra they generate has a nonzero central charge, which equals the dimension of the target space $M$. Therefore, to get a truly superconformal theory, one has to manage to cancel that central charge. This is achieved in topological string theory we will discuss later via topological twisting.

There is another class of (2,2) rigidly supersymmetric Lagrangians that are usually used called superpotentials. The construction of these terms involving integrating over ``half superspace''. Let's now make this concrete using supermanifolds.

Recall that we have projections from an $\mathcal{N}=2$ super Riemann surface to a pair of complex supermanifolds of dimension $1|1$. Here we have a similar set of projections
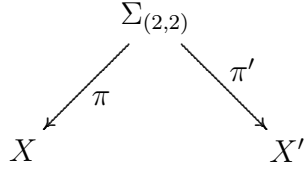
\begin{equation*}
\begindc{\commdiag}[50]
\obj(1,1)[s]{$\Sigma_{(2,2)}$}
\obj(0,0)[x]{$X$}
\obj(2,0)[x']{$X'$}
\mor{s}{x}{$\pi$}
\mor{s}{x'}{$\pi'$}
\enddc
\end{equation*}
where $X$ and $X'$ are smooth supermanifold of dimension $2|2$, with local coordinates $(\widetilde{z}, z|\widetilde{\theta}^+, \theta^+)$ and $(\widetilde{z}, z|\widetilde{\theta}^-, \theta^-)$, respectively. The reason for the existence of these projections is simple: $\Sigma_{(2,2)}$ is split. These projections can be understood as setting the appropriate odd coordinates to 0. Then it can be shown that a chiral superfield $\Phi$ is the pullback of a complex function $\varphi: X \rightarrow \mathbb{C}$, while an antichiral superfield is the pullback of a complex function $\varphi': X' \rightarrow \mathbb{C}$. 

Now, the superpotential terms are given by
\begin{equation}
\int_{X} \mathcal{D}(\widetilde{z},z|\widetilde{\theta}^+, \theta^+) W(\varphi^i) + \int_{X'} \mathcal{D}(\widetilde{z},z|\widetilde{\theta}^-, \theta^-) W'(\varphi'^i).
\end{equation}
Once again, we are not allowed to assume $W'$ to be the complex conjugate of $W$. Note this action is not defined by integration over the entire $\Sigma_{(2,2)}$, so it feels a bit bizarre. However, first notice that $\Sigma_{(2,2)}$, $X$ and $X'$ all have the same reduced space, on which usual Lagrangians with component fields are usually defined. Furthermore, since $\Sigma_{(2,2)}$ is split, we can integrate out the odd coordinates and obtain an ordinary Lagrangian in terms of component fields with integration over that common reduced space, which is what's physically important. That said, any sort of potential term will ruin conformal symmetry, thus we will not discuss these terms in this paper.

There is another class of (2,2) superfields called twisted chiral superfields. A twisted chiral superfield is a complex function $\Omega: \Sigma_{(2,2)} \to \mathbb{C}$, satisfying
\begin{equation}
D_- \Omega = \widetilde{D}_+ \Omega = 0.
\end{equation}
Similarly, an twisted antichiral superfield is a complex function $\Omega': \Sigma_{(2,2)} \rightarrow \mathbb{C}$ satisfying
\begin{equation}
D_+ \Omega' = \widetilde{D}_- \Omega' = 0.
\end{equation}
One can formulate Lagrangians for twisted superfields, completely analogous to the case of chiral superfields. We will not discuss this further.

\section{Topological String Worldsheet}

\subsection{Semirigid Super Riemann Surface} \label{ssrs}
\subsubsection{The Stage}
What we have discussed so far are theories that are usually called physical, in the sense that they are typical quantum field theories with complicate dynamics. With supersymmetry, one can derive many useful information about these physical theories, but in general it is really hard to understand them fully. In contrast, topological field theories have much smaller Hilbert spaces (which are in fact finite-dimensional), with basically all of their properties derived exactly. Let's now discuss how to obtain topological field theories from physical theories we have discussed.

Let's start with an $\mathcal{N}=2$ super Riemann surface $S$, with distinguished odd line bundles $\mathcal{D}_-$ and $\mathcal{D}_+$. Then by topological twist we mean the following procedure \cite{distler-nelson}: we take one of these two odd line bundles, say $\mathcal{D}_-$, to be trivial, with a global section $D_-$ given. 

To see what is happening, let's recall the superconformal coordinate transformations on $S$ from equation (\ref{superconformal-transf}):
\begin{equation} \label{superconformal-transf-again}
\begin{split}
z' & = f(z) + \theta^- a(z) \rho(z) + \theta^+ e(z) \alpha(z) + \theta^- \theta^+ \partial_{z} (\alpha(z) \rho(z)), \\
{\theta^-}^{\prime} & = \alpha(z) + \theta^- a(z) + \theta^- \theta^+ \partial_z \alpha(z), \\
{\theta^+}^{\prime} & = \rho(z) + \theta^+ e(z) - \theta^- \theta^+ \partial_z \rho(z), 
\end{split}
\end{equation}
with the following constraint
\begin{equation}
e a = \partial_z f + \alpha \partial_z \rho + \rho \partial_z \alpha,
\end{equation}
where $f,\ a, \ e$ are even functions of $z$, while $\alpha, \ \rho$ are odd functions of $z$. 

Now by taking $\mathcal{D}_-$ to be trivial with a global section $D_-$ given, we mean that any other section of $\mathcal{D}_-$ should be equal to $D_-$. The above superconformal transformation (\ref{superconformal-transf-again}) induces transformations on sections of $\mathcal{D}_-$:
\begin{equation}
D_- = (D_- {\theta^-}^{\prime}) D_-',
\end{equation}
where
\begin{equation}
D_-' = \frac{\partial}{\partial {\theta^-}^{\prime}} + {\theta^+}^{\prime} \frac{\partial}{\partial z'}
\end{equation}
is another section of $\mathcal{D}_-$. This means that we need $D_- {\theta^-}^{\prime} = 1$, which leads to $\alpha = \text{constant}$ and $a = 1$. Notice now we are allowed to assign spin 0 to $\theta^-$ and spin 1 to $\theta^+$, but still maintain their anti-commuting nature. Let's futher constraint $\alpha = 0$. This is the $-$ twist introduced in \cite{witten-tft}. In contrast, by demanding $\mathcal{D}_+$ to be trivial with a global section given, we obtain the $+$ twist in \cite{witten-tft}. The resulting supermanifolds are called semirigid super Riemann surfaces \cite{distler-nelson}, denoted as $\mathbf{S}^{\mp}$. Obviously $\mathbf{S}^- \cong \mathbf{S}^+$, so we will sometimes just omit the superscript to avoid clustering of indices.

More precisely, with the $-$ twist, $\theta^-$ is now an anticommuting section of the trivial line bundle $\mathcal{O} \to \mathbf{S}_{\text{red}}$, while $\theta^+$ becomes a (1,0) form on $\mathbf{S}_{\text{red}}$. The exact opposite situation happens for the $+$ twist.

With the $-$ twist, the superconformal coordinate transformations (\ref{superconformal-transf-again}) become
\begin{equation} \label{semirigid-transf}
\begin{split}
z' & = f(z) + \theta^-  \rho(z), \\
{\theta^-}^{\prime} & = \theta^-,   \\
{\theta^+}^{\prime} & = \rho(z) + \theta^+ \partial_z f(z) - \theta^- \theta^+ \partial_z \rho(z). 
\end{split}
\end{equation}
We call these transformations \textit{semirigid coordinate transformations}. We see explicitly that $\theta^-$ transforms trivially. Therefore, our assignment of spin 0 to $\theta^-$ is valid.

Recall from section \ref{srs} that we have canonical projections from a $\mathcal{N} = 2$ super Riemann surface to a couple of complex supermanifolds with dimension $1|1$. We can do the exact same thing to our semirigid super Riemann surface $\mathbf{S}$, although here it is even simpler. In particular, the projection
\begin{equation}
\mathbf{S}^- \to X^-
\end{equation}
produce the following coordinate transformations on $X^-$ (with local coordinates $(w|\theta^-)$ where $w = z + \theta^-\theta^+$):
\begin{equation}
\begin{split}
w' & = f(w) + \theta^-  2\rho(w), \\
{\theta^-}^{\prime} & = \theta^-.  
\end{split}
\end{equation}
Note that exactly like the physical $\mathcal{N}=2$ case, here the data from the coordinate transformation functions $f(w)$ and $\rho(w)$ on $X^-$ uniquely determine the semirigid coordinate transformation on the upstairs semirigid super Rieman surface $\mathbf{S}^-$.

On the other hand, we also have a analogous projection
\begin{equation}
\mathbf{S}^- \to X^+
\end{equation}
where $X^+$ has local coordinates $(u|\theta^+)$, with $u = z - \theta^-\theta^+$, that transform as
\begin{equation}
\begin{split}
u' & = f(u), \\
{\theta^+}^{\prime} & = \rho(u) + \theta^+ \partial_u f(u).
\end{split}
\end{equation}
Clearly we see that $X^+$ is globally isomorphic to $\mathit{\Pi} TX^+_{\text{red}}$, the tangent bundle of the reduced space $X^+_{\text{red}}$ with its statistics reversed along the fibers. Once again, the data from the coordinate transformation functions $f(u)$ and $\rho(u)$ on $X^+$ uniquely determine the semirigid coordinate transformation on the upstairs semirigid super Rieman surface $\mathbf{S}^-$.

At this point it is natural to remark that the supermoduli space of $\mathbf{S}^-$ (as a semirigid surface) is isomorphic to the supermoduli space of $X$ (or $X'$) as a generic complex supermanifold of dimension $1|1$. This is very useful when one's purpose is to study the properties of this kind of supermoduli spaces, such as when proving a generic supermoduli space of this sort is not split.

\subsubsection{Semirigid Vector Fields}

How do we understand these topological twists in terms of actions of superconformal vector fields in equation (\ref{superconformal-vector})? Let's take the $-$ twist as an example (the $+$ twist is completely analogous). Now that we have a trivial $\mathcal{D}_-$ with a global section $D_-$ given, we must require that the infinitesimal coordinate transformations generated by these vector fields to vanish. In other words, they should commute/anticommute with $D_-$. Then from (\ref{superconformal-vector-act-D}) we get
\begin{equation} \label{topological-vector-act-D}
\begin{split}
\partial_z g - \theta^-\theta^+ \partial_z\partial_z g  &= 0, \\
k - \theta^-\theta^+ \partial_z k  &= 0, \\
\theta^+ \partial_z \alpha^-  &= 0.
\end{split}
\end{equation}
The last equation can only be solved by $\alpha^- = c$ where $c$ is a constant. Then we see $\mathcal{G}^- = c \mathcal{Q}_-$. However, now we have three constraint on four functions, so we are only left with one degree of freedom, which is not what we want (we have two remaining functions $f$ and $\rho$ from coordinate transformations). This is saying that the original $\mathcal{N}=2$ superconformal vector fields are not what we need on a semirigid super Riemann surface.

We can achieve a bypass to the rest of these equations while maintaining the correct number of independent generators by the following ``twisting'':
\begin{equation}
\mathcal{T} \mapsto \mathcal{T}^- = \mathcal{T} - \frac{1}{2} \partial_z \mathcal{J} = g \partial_z + \frac{1}{2} \partial_z (g-k) \theta^- \partial_{\theta^-} + \frac{1}{2} \partial_z (g+k) \theta^+ \partial_{\theta^+},
\end{equation}
together with the constraint $g(z) - k(z) = \text{constant}$. Then one can show that
\begin{equation}
[\mathcal{T}^-, D_-] = -\frac{1}{2} \left( \partial_z (g-k) - \theta^-\theta^+ \partial_z\partial_z (g-k) \right) D_{-} = 0.
\end{equation}
So with this new vector fields $\mathcal{T}^-$, we see that the given global section $D_-$ is preserved. We call the resulting vector fields $\mathcal{T}^-, \mathcal{J}, \mathcal{Q}_-$, and $\mathcal{G}^+$ \textit{semirigid vector fields}. Note that now $\mathcal{D}_-$ is trivial, we see $\mathcal{G}^- = c \mathcal{Q}_-$ can have spin 0. We can count that now we have one independent even generator and one independent odd generator left, precisely corresponding to the remaining functions $f$ and $\rho$.

Completely analogously, the $+$ twist can be achieved by taking $\mathcal{G}^+ = c \mathcal{Q}_+$ for a constant $c$, together with
\begin{equation}
\mathcal{T} \mapsto \mathcal{T}^+ = \mathcal{T} + \frac{1}{2} \partial_z \mathcal{J} = g \partial_z + \frac{1}{2} \partial_z (g+k) \theta^- \partial_{\theta^-} + \frac{1}{2} \partial_z (g-k) \theta^+ \partial_{\theta^+}, 
\end{equation}
and we impose the same constraint $g(z) - k(z) = \text{constant}$ as the above case of the $-$ twist. Here we see
\begin{equation}
[\mathcal{T}^+, D_+] = -\frac{1}{2} \left( \partial_z (g-k) + \theta^-\theta^+ \partial_z\partial_z (g-k) \right) D_{+} = 0.
\end{equation}

\subsubsection{Topological Field Theory}

Now let's construct the Lagrangian for our topological field theories that live on topological worldsheets. Notice that the Berezinian line bundle of $\mathbf{S}$, $\text{Ber}(\mathbf{S})$, is still a canonically trivial line bundle, just as in the case of $\mathcal{N}=2$ super Riemann surfaces. Therefore, all of our discussions in section \ref{22} about constructing Lagrangians have direct analogues here. 

However, there is one major difference here: we don't have any spinors in our twisted theory. As such, we can demand reality conditions on the anticommuting coordinates. Therefore we don't actually need to construct a middle-dimensional cycle to define our worldsheet as we did in the physical cases. Nonetheless, if we want to distinguish left-moving and right-moving degree of freedom, it is still rather convenient to use a similar approach (albeit trivially): we take the worldsheet of topological strings to be a smooth supermanifold $\Sigma_T$ diagonally embedded as 
\begin{equation}
\Sigma_T \hookrightarrow \mathbf{S}_L \times \mathbf{S}_R,
\end{equation}
where $\mathbf{S}_L$ and $\mathbf{S}_R$ are both semirigid super Riemann surfaces, characterizing left-moving and right-moving degrees of freedom, respectively. As usual, we parametrize $\mathbf{S}_L$ with local superconformal coordinates $(\widetilde{z}|\widetilde{\theta}^-,\widetilde{\theta}^+)$, and $\mathbf{S}_R$ with local superconformal coordinates $(z|\theta^-,\theta^+)$. This will be handy when we want to distinguish the two usual different types of topological string worldsheets which will be defined next.

If $\mathbf{S}_L$ is endowed with the $-$ twist and $\mathbf{S}_R$ is endowed with the $+$ twist, we call such a worldsheet is of \textit{A-type}, denoted as $\Sigma_T^A$. In this case, $\theta^+$ and $\widetilde{\theta}^-$ have spin 0, i.e. they are (anticommuting) sections of the trivial line bundle $\mathcal{O} \to {\Sigma_T^A}_{\text{red}}$; while $\theta^-$ and $\widetilde{\theta}^+$ both have spin 1, i.e. $\theta^-$ is a section of the holomorphic cotangent bundle ${T^*}^{1,0}{\Sigma_T^A}_{\text{red}}$, while $\theta^+$ is a section of the antiholomorphic cotangent bundle ${T^*}^{0,1}{\Sigma_T^A}_{\text{red}}$. In other words, $\theta^-$ is a (1,0) form on ${\Sigma_T^A}_{\text{red}}$ while $\widetilde{\theta}^+$ is a (0,1) form on ${\Sigma_T^A}_{\text{red}}$. In this case, the reality condition we can impose on local coordinates is that $(z | \theta^-, \theta^+)$ is the complex conjugate of $(\widetilde{z} | \widetilde{\theta}^+, \widetilde{\theta}^-)$ (note the order). As such, $\mathcal{Q}_+$ and $\widetilde{\mathcal{Q}}_-$ are complex conjugate to each other.

In comparison, if both $\mathbf{S}_L$ and $\mathbf{S}_R$ are endowed with the $-$ twist, we call such a worldsheet is of \textit{B-type}, denoted as $\Sigma_T^B$. In this case, $\theta^-$ and $\widetilde{\theta}^-$ have spin 0, i.e. they are anticommuting scalars. While $\theta^+$ and $\widetilde{\theta}^+$ have spin 1. In this case, the reality condition we can impose on those anticommuting coordinates is that $(z | \theta^-, \theta^+)$ is the complex conjugate of $(\widetilde{z} | \widetilde{\theta}^-, \widetilde{\theta}^+)$.

Now we are ready to write down the action for the A (or B) model topological field theory:
\begin{equation} \label{tft-action}
I_{\text{TFT}} = \int_{\Sigma_T} \mathcal{D}(\widetilde{z},z|\widetilde{\theta}^-,\widetilde{\theta}^+, \theta^-,\theta^+) K(\widetilde{\Phi}^{\bar{\imath}}, \Phi^i),
\end{equation}
where $\widetilde{\Phi}^{\bar{\imath}}$ and $\Phi^i$ are antichiral and chiral superfields as defined in section \ref{22} with appropriate twists. The target space $M$ is a Calabi-Yau manifold, as we explained in section \ref{22}. Let's take the A mode as an example. Recall we have expansion
\begin{equation}
\begin{split}
\Phi^i  =& \ \phi^i + \theta^+ \psi^i + \widetilde{\theta}^+ \lambda^i + \theta^+ \widetilde{\theta}^+ F^i - \theta^-\theta^+ \partial_z \phi^i - \widetilde{\theta}^- \widetilde{\theta}^+ \partial_{\widetilde{z}}\phi^i \\
& - \theta^+ \widetilde{\theta}^- \widetilde{\theta}^+ \partial_{\widetilde{z}} \psi^i - \theta^- \theta^+ \widetilde{\theta}^+ \partial_z\lambda^i + \theta^- \theta^+ \widetilde{\theta}^- \widetilde{\theta}^+ \partial_z \partial_{\widetilde{z}} \phi^i, \\
\widetilde{\Phi}^{\bar{\imath}} =& \ \widetilde{\phi}^{\bar{\imath}} + \theta^- \psi'^{\bar{\imath}} + \widetilde{\theta}^- \lambda'^{\bar{\imath}} +  \theta^- \widetilde{\theta}^- F'^{\bar{\imath}}  + \theta^-\theta^+ \partial_{\widetilde{z}} \widetilde{\phi}^{\bar{\imath}} + \widetilde{\theta}^- \widetilde{\theta}^+ \partial_z\widetilde{\phi}^{\bar{\imath}} \\
&+ \theta^- \widetilde{\theta}^- \widetilde{\theta}^+ \partial_z \psi'^{\bar{\imath}} + \theta^- \theta^+ \widetilde{\theta}^- \partial_{\widetilde{z}}\lambda'^{\bar{\imath}} + \theta^- \theta^+ \widetilde{\theta}^- \widetilde{\theta}^+ \partial_z \partial_{\widetilde{z}} \widetilde{\phi}^{\bar{\imath}},
\end{split}
\end{equation}
Now that $\theta^+$ and $\widetilde{\theta}^-$ are anticommuting scalars, we see that $\psi^i$ and $\lambda'^{\bar{\imath}}$ are anticommuting scalars, too. In contrast, $\psi'^{\bar{\imath}}$ is now a (1,0) form, while $\lambda^i $ is a (0,1) form, both on ${\Sigma_T^A}_{\text{red}}$. In other words, the fermionic fields in the physical (2,2) theory become ghost fields in the topological theory after twisting.

Recall that A model can be defined on any almost complex manifold. In contrast, anomaly cancellation conditions require $M$ to be a Calabi-Yau manifold in order for the B model to be well defined. In our case this is always true because we always take $M$ to be Calabi-Yau.

Note we have not coupled our topological field theories to worldsheet topological gravity yet, so we are not dealing with full topological string theories yet. We will do that later, after we discuss the supermoduli spaces\footnote{To be rigorous mathematically, these objects should be called supermoduli stacks. However, we will not go into details along these lines.} of semirigid super Riemann surfaces.

\subsection{Supermoduli Space} \label{moduli}

Now let's analyze the supermoduli spaces of semirigid super Riemann surfaces. Our discussion will follow the discussion in \cite{witten-srs} on $\mathcal{N} = 1$ super Riemann surfaces. Let $\mathbf{S}$ be a semirigid super Riemann surface, with an open cover $\{ U_{\alpha} \}$. Then the local data says that each $U_{\alpha}$ is an open subset of $\mathbb{C}^{1|2}$, and the entire $\mathbf{S}$ is constructed by gluing all these $U_{\alpha}$'s together by semirigid coordinate transformations (\ref{semirigid-transf}). To the first order, the gluing data on the overlap $U_{\alpha} \cap U_{\beta}$ can be deformed by topologcial vector fields $\phi_{\alpha\beta}$. As such, one need the following cocycle condition:
\begin{equation}
\phi_{\alpha\beta} + \phi_{\beta\gamma} + \phi_{\gamma\alpha} = 0,
\end{equation}
on every triple overlap $U_{\alpha} \cap U_{\beta} \cap U_{\gamma}$. In particular, this says that $\phi_{\alpha\beta}$ is a one-cocycle.

On the other hand, to have consistent data on all overlaps, we also need the following equivalence relation:
\begin{equation}
\phi_{\alpha\beta} \sim \phi_{\alpha\beta}' = \phi_{\alpha\beta} + \phi_{\alpha} - \phi_{\beta},
\end{equation}
on any $U_{\alpha} \cap U_{\beta}$. This means that we are defining a family of one-cocycles $\phi_{\alpha\beta}$ that are equivalent to each other up to coboundaries. Therefore, the first order deformations of a semirigid super Riemann surface is given by elements of a certain sheaf cohomology group. 

Let $\mathcal{V}$ be the sheaf of semirigid vector fields over a semirigid super Riemann surface $\mathbf{S}$. Then the sheaf cohomology group we want is $H^1(\mathbf{S}, \mathcal{V})$. Now we understand the elements of this cohomology group parametrize first order deformations of $\mathbf{S}$, or in other words, the local data on the supermoduli space $\mathfrak{M}$ of semirigid super Riemann surfaces. Let $T\mathfrak{M}_{\mathbf{S}}$ be the tangent space of $\mathfrak{M}$ at a point corresponding to $\mathbf{S}$. It should be clear now that $T\mathfrak{M}_{\mathbf{S}} = H^1(\mathbf{S}, \mathcal{V})$\footnote{An alternative way to obtain this result using deformation theory can be found in appendix \ref{deformation}.}.

If $\mathbf{S}$ is split, then we can do more. This corresponds to taking $\rho = 0$ in semirigid coordinate transformations (\ref{semirigid-transf}). Then we see that $\theta^-$ becomes a section of the trivial bundle $\mathcal{O}$ over the reduced space $\mathbf{S}_{\text{red}}$, which is an ordinary Riemann surface. We also see that $\theta^+$ transform exactly like $dz$, so it is a section of the canonical bundle $K$ of $\mathbf{S}_{\text{red}}$. Therefore, we see that our local mode $V \to \mathbf{S}_{\text{red}}$ is simply given by 
\begin{equation}
V \cong \mathcal{O} \oplus K^{-1} \cong \mathcal{O} \oplus T\mathbf{S}_{\text{red}},
\end{equation}
where $T\mathbf{S}_{\text{red}}$ is the tangent bundle of the reduced space $\mathbf{S}_{\text{red}}$.

In this case, we have a natural decomposition 
\begin{equation}
\mathcal{V} = \mathcal{V}_+ \oplus \mathcal{V}_-,
\end{equation}
where $\mathcal{V}_+$ and $\mathcal{V}_-$ are the even part and odd part of $\mathcal{V}$, respectively. Then we have
\begin{equation}
H^1(\mathbf{S},\mathcal{V}) = H^1(\mathbf{S},\mathcal{V}_+) \oplus H^1(\mathbf{S},\mathcal{V}_-).
\end{equation}
Observe that, from the expression of $\mathcal{N}=2$ superconformal vector fields in (\ref{superconformal-vector}) with appropriate topological twists. More explicitly, from the expressions of $\mathcal{T}$ and $\mathcal{J}$, we see that $\mathcal{V}_+$ is generated by $\partial_z$, while from the expressions of $\mathcal{G}^{\pm}$ we see that $\mathcal{V}_-$ is generated by $\partial_{\theta^{\pm}}$. Therefore, we have
\begin{equation}
\begin{split}
\mathcal{V}_+ &\cong T\mathbf{S}_{\text{red}}, \\
\mathcal{V}_- &\cong  V \cong \mathcal{O} \oplus T\mathbf{S}_{\text{red}}.
\end{split}
\end{equation}

Let's apply Riemann-Roch theorem for a line bundle $L$ over $\mathbf{S}_{\text{red}}$ with degree $n$:
\begin{equation}
\text{dim} \ H^0(\mathbf{S}_{\text{red}}, L) - \text{dim} \ H^1(\mathbf{S}_{\text{red}}, L) = 1 - g + n,
\end{equation}
where $g$ is the genus of $\mathbf{S}_{\text{red}}$. Notice that $H^0(\mathbf{S}_{\text{red}}, L) = 0$. Now we take $L$ to be $T\mathbf{S}_{\text{red}}$. When $g \ge 2$, we have $n = 2 - 2g < 0$. Therefore we see $\text{dim} \ H^1(\mathbf{S}_{\text{red}}, T\mathbf{S}_{\text{red}}) = 3g - 3$. 

We conclude that
\begin{equation}
\text{dim} \ \mathfrak{M}_g = 3g - 3 | 3g - 3, \ g \ge 2,
\end{equation}
which is the correct dimension for topological string theory. In general, we have to take into consideration the automorphisms of $\mathbf{S}$. Let $\mathbf{G}$ be the supergroup of automorphisms of $\mathbf{S}$. Then we have
\begin{equation}
\text{dim} \ \mathfrak{M}_g - \text{dim} \ \mathbf{G} = 3g - 3 | 3g - 3, \ g \ge 2.
\end{equation}

In general, we would not expect the moduli space $\mathfrak{M}_g$ of genus $g$ semirigid super Riemann surfaces to be split. In fact, we wouldn't expect it to be holomorphically projected generically. This is part of our motivation to revisit topological string theory, so that we can formulate everything naturally in terms of super manifolds and supermoduli. The proof of the nonsplitness of these supermoduli spaces will appear in a companion paper \cite{distler-jia-2}.

Strictly speaking, we need to worry about the fact that $\mathfrak{M}_{g}$ is noncompact. There is a natural way to compactify it, analogouse to the Deligne-Mumford compactifications of $\mathcal{M}_{g,N}$ and the supermoduli space of $\mathcal{N}=1$ super Riemann surfaces used in physical string theory. Let's denote the compactified spaces as $\mathfrak{\overline{M}}_{g}$. Everything we will discuss should be defined on this compact space.

Another important property of $\mathfrak{M}_{g}$ is that its Berezinian line bundle $\text{Ber}(\mathfrak{M}_{g})$ (thinking of $\mathfrak{M}_{g}$ as a smooth supermanifold) is trivial. We prove this in appendix \ref{berezinian}. Later when we define topological string amplitudes as integrals over $\mathfrak{M}_{g}$, we will need to worldsheet field theory pathintegral to provide a section of $\text{Ber}(\mathfrak{M}_{g})$, i.e. a function on $\mathfrak{M}_{g}$.

\section{Topological Strings}

\subsection{Coupling to Topological Gravity}

So far we have discussed pure field theory which doesn't involve gravity. Put another way, we have ignoored couplings to worldsheet gravity. To obtain topological string theory, morally one should couple our A/B model topological field theory to some appropriate worldsheet gravity \cite{witten-ts, dvv}. In particular, this worldsheet gravity theory should enjoy a twisted $\mathcal{N}=2$ symmetry, just as the matter sector. The suitable version of worldsheet gravity is called two-dimensional topological gravity. As argued in \cite{distler-nelson}, this theory can be viewed as a theory obtained by twisting $\mathcal{N}=2$ sperconformal gravity, thus can be easily coupled to our topological field theory.\footnote{There is another version of topological string theory defined on Calabi-Yau three-folds, using the analogy between topologcial A/B models with bosonic string theory. In this case one doesn't introduce additional ghost fields. We will not discuss this version in this paper.} 

Morally this means we need to go through the usual gauge-fixing and introducing the superconformal ghost fields, or superghosts, in our theory. As usual, the holomorphic superghosts are given by a pair of complex superfields $C, B$ on $\Sigma_T^A$,\footnote{We will focus on the A-model topological string theory in the rest of this paper. The discussion of B-model is analogous.} satisfying
\begin{equation}
\begin{split}
& D_- C = a,  \\
& \widetilde{D}_{\mp} C =  \widetilde{D}_{\mp} B = 0,
\end{split}
\end{equation}
where $a \in \mathbb{C}$ is a complex constant. The first condition is precisely equivalent to the $+$ twist we defined earlier. The conditions on the second line are saying the $B$ and $C$ are holomorphic quantities on $\Sigma_T^A$, in the sense defined in appendix \ref{deformation}.

Geometrically, the superghost $C$ is, as always, the same as superconformal superfield $\mathcal{Y}(z|\theta^{\mp})$ we defined in (\ref{superconformal-vector}) (with spin-statistics reversed). So from appendix \ref{berezinian} we know that the superghost $C$ must be a (anticommuting) section of $\mathit{\Pi} (\mathcal{D}_- \otimes \mathcal{D}_+)$ (pulled-back to $\Sigma_T^A$). As such, we see that the superghost $B$ is a (anticommuting) section of $\mathit{\Pi} (\mathcal{D}_- \otimes \mathcal{D}_+)^{-1}$ (pulled-back to $\Sigma_T^A$), because the Berezinian line bundle of $\Sigma_T^A$ is trivial.

Similarly, the antiholomorphic superghosts are given by a pair of complex superfields $\widetilde{C}, \widetilde{B}$ on $\Sigma_T^A$, satisfying
\begin{equation}
\begin{split}
& \widetilde{D}_+ \widetilde{C} = \widetilde{a},  \\
& D_{\pm} \widetilde{C} =  D_{\pm} \widetilde{B} = 0,
\end{split}
\end{equation}
where $\widetilde{a} \in \mathbb{C}$ is a complex constant (which is simply the complex conjugate of $a$ above). Therefore, $\widetilde{B}$ and $\widetilde{C}$ are antiholomorphic quantities on $\Sigma_T^A$. Geometrically, from the above discussion we see that the superghost $\widetilde{C}$ must be a section of $\mathit{\Pi} (\widetilde{\mathcal{D}}_- \otimes \widetilde{\mathcal{D}}_+)$ (pulled-back to $\Sigma_T^A$). As such, again we see that the superghost $\widetilde{B}$ is a section of $\mathit{\Pi} (\widetilde{\mathcal{D}}_- \otimes \widetilde{\mathcal{D}}_+)^{-1}$ (pulled-back to $\Sigma_T^A$). 

What about component fields? In a physical $(2,2)$ superconformal gravity (defined on a general  $\mathcal{N}=2$ super Riemann surface), we have the usual expansion
\begin{equation} 
\begin{split}
C & = c^z + \theta^- \gamma^+ + \theta^+ \gamma^- + \theta^- \theta^+  c'^z, \\
B & = b'_z + \theta^- \beta_{-z} - \theta^+ \beta_{+z} - \theta^-\theta^+ (b_{zz} + \partial_z b'_z),
\end{split}
\end{equation}
where we have the usual superconformal ghosts $(b_{zz}, c^z)$ and a pair of $(\beta_{\pm}, \gamma^{\pm})$. In addition, since  there is a U(1) gauge filed in (2,2) supergravity, we also have additional U(1) ghost fields $(b'_z, c'^z)$.

On our semirigid super Riemann surface $\mathbf{S}_R$, things are a bit different after twisting \cite{distler-nelson}. Let's focus on the A twist, which results in
\begin{equation} \label{ghost}
\begin{split}
C & = c^z + \theta^- a + \theta^+ \gamma^- + \theta^- \theta^+ \partial_z c^z, \\
B & = b'_z + \theta^- \beta_{-z} - \theta^+ \beta_{+z} - \theta^-\theta^+ (b_{zz} + \partial_z b'_z),
\end{split}
\end{equation}
where $a$ is the constant in the definition of superghosts. As we will see later \cite{distler-nelson}, ghost fields $b'_z$ and $\beta_{+z}$ will not contribute to our semirigid physics; they don't show up in ghost Lagrangian nor do they show up in physical observalbes. So to simplify notation, let's write $\beta_z = \beta_{-z}$ and $\gamma = \gamma^-$.

Similarly, on $\mathbf{S}_L$ we introduce the antiholomorphic superghost and antisuperghost
\begin{equation}
\begin{split}
\widetilde{C} & = \widetilde{c}^{\widetilde{z}} + \widetilde{\theta}^- \widetilde{\gamma}^+ + \widetilde{\theta}^+ \widetilde{a} + \widetilde{\theta}^- \widetilde{\theta}^+ \partial_{\widetilde{z}} \widetilde{c}^{\widetilde{z}}, \\
\widetilde{B} & = \widetilde{b}'_{\widetilde{z}} + \widetilde{\theta}^- \widetilde{\beta}_{-\widetilde{z}} - \widetilde{\theta}^+ \widetilde{\beta}_{+\widetilde{z}} - \widetilde{\theta}^-\widetilde{\theta}^+ (\widetilde{b}_{\widetilde{z}\widetilde{z}} + \partial_{\widetilde{z}} \widetilde{b}'_{\widetilde{z}}),
\end{split}
\end{equation}
where $\widetilde{a}$ is the complex conjugate of $a$. Again, similarly, ghost fields $\widetilde{b}'_{\widetilde{z}}$ and $\widetilde{\beta}_{-\widetilde{z}}$ will not contribute to our semirigid physics. So we write $\widetilde{\beta}_{\widetilde{z}} = \widetilde{\beta}_{+\widetilde{z}}$ and $\widetilde{\gamma} = \widetilde{\gamma}^+$.

In terms of ordinary geometry on the reduced space of our topological string worldsheet, we see that $\gamma$ and $\widetilde{\gamma}$ are simply functions (sections of the trivial bundle $\mathcal{O}$), while $\beta_z$ is a (commuting) $(1,0)$ form and $\widetilde{\beta}_{\widetilde{z}}$ is a (commuting) $(0,1)$ form.

Now we are ready to write down the action for our superghosts:
\begin{equation}
I_{\text{ghost}} = \int_{X} \mathcal{D}(\widetilde{z},z|\theta^-,\theta^+)  ( B \partial_{\widetilde{z}} C ) + \int_{\widetilde{X}} \mathcal{D}(\widetilde{z},z|\widetilde{\theta}^-, \widetilde{\theta}^+) ( \widetilde{B} \partial_z \widetilde{C} ).
\end{equation} 
It is important to observe that those extra ghosts ($\widetilde{b}'_{\widetilde{z}}, \widetilde{\beta}_{-\widetilde{z}}, b'_z, \beta_{+z})$ don't appear in our ghost action, which is exactly what we expected. Note that we are not integrating over the entire $\Sigma_T^A$. Rather, we are only integrating over submanifolds $X$ (parametrized by local coordinates $(\widetilde{z},z|\theta^-,\theta^+)$) and $\widetilde{X}$ (parametrized by local coordinates $(\widetilde{z},z|\widetilde{\theta}^-, \widetilde{\theta}^+)$). This is similar to the situation in section \ref{22} where we constructed the action for (2,2) superpotential. 

Note that this action in terms of component fields ($b, c, \beta, \gamma$) formally looks exactly the same as in the physical superstring case. However, the conformal dimension of both $b$ and $\beta$ are 2, which is a consequence of topological twist. As such, the total central charge from this ghost sector is 0, which is needed since the matter sector has a twisted $\mathcal{N} = 2$ algebra which has central charge 0. Therefore, we have
\begin{equation}
\begin{split}
T_G & =  - 2 b \partial c + c \partial b - 2 \beta \partial \gamma - \gamma \partial \beta, \\
J_G & =  b c - 2 \beta \gamma,\\
Q_G & =  b \gamma, \\
G_G & = - 2 \beta \partial c - c \partial\beta - b,
\end{split}
\end{equation}
which are again field theory representations of our semirigid vector fields so they generate the twisted $\mathcal{N} = 2$ algebra. In particular, $Q_G$ (and so $\widetilde{Q}_G$) generates the needed BRST symmetry to define a cohomological topological theory. As such, $Q_G$ ($\widetilde{Q}_G$) is the representation of $\mathcal{Q}_+$ ($\widetilde{\mathcal{Q}}_-$) on the Hilbert space.

We should remark that we have a rather peculiar expression for our ghost supercurrent $G_G$, in the sense that it is inhomogeneous in terms of ghost numbers from $bc$ system and $\beta\gamma$ system because of the last term $b$ in it. We will see the consequence of this in the next section when we write down topological string amplitudes. 

Now we are ready to define the action for the A model topological string theory as
\begin{equation}
I =  I_{\text{TFT}} + I_{\text{ghost}},
\end{equation}
where we added the action for superghosts. It is worth pointing out that we have two independent ghost systems here: the $B$-$C$ superghosts introduced in this section, as well as the ``ghosts'' (twisted fermions) from topological field theory.

However, in order to define a topological string theory, it is not enough to just have a Lagrangian. We need to choose a set of observables and obtain correlation functions between them. Geometrically, we will need to define punctures on our seimirigid super Riemann surfaces. Let's get to it now.

\subsection{Punctures and Observables}

In physical superstring theory, perturbative correlation functions between physical observables are calculated by first inserting vertex operators on string worldsheets. To insert observables on our worldsheets of topological strings, we need the notion of punctures. On a ordinary Riemann surface, a puncture is simply a fixed point the that surface, in other words a divisor. On a semirigid super Riemann surfaces with the $-$ twist, a puncture is specified by $(z_0|\theta^+_0)$, because $\theta^-$ is globally a constant. In other words, we have a divisor of the form $z - z_0 - \theta^+ \theta^+_0$. Therefore, adding a puncture increases the dimension of the supermoduli space by $1|1$. This is similar to the Neveu-Schwarz (NS) punctures in superstring theory. Therefore,
\begin{equation}
\text{dim} \ \mathfrak{M}_{g,N} - \text{dim} \ \mathbf{G} = 3g - 3 + N | 3g - 3 + N, \ g \ge 2,
\end{equation}
where $\mathfrak{M}_{g,N}$ denotes the supermoduli space of semirigid super Riemann surfaces of genus $g$ with $N$ punctures. Let's denote $K = 3g - 3 +N$ for later convenience. 

In defining a topological string theory, we have the ability to select a set of observables that have expected properties and we define them as our physical observables. As usually done in topological theories, this is achieved by taking the cohomology of a nilpotent symmetry charge. Here we have a natural candidate for such a charge, namely the BRST charge in our theory. To see this, we define the following vector field on our topological string world sheet $\Sigma_T^A$:
\begin{equation}
\mathcal{Q} = \widetilde{\mathcal{Q}}_- + \mathcal{Q}_+,
\end{equation}
where we used the fact from section \ref{ssrs} that in the A model topological field theory, we have $\widetilde{\mathcal{G}}^- = \widetilde{\alpha}^- \widetilde{\mathcal{Q}}_-$ and $\mathcal{G}^+ = \alpha^+ \mathcal{Q}_+$. We further constrain $\widetilde{\alpha}^- =  \alpha^+ = \alpha$ which is a scalar constant. One can easily check that $\mathcal{Q}^2 = 0$. Let's call the representation of $\mathcal{Q}$ on the Hilbert space as $Q_S$. Then it follows that $Q_S$ is a scalar with $Q_S^2 = 0$. On the other hand, we also have a usual BRST charge $Q_{\text{V}}$ from the Virasoro generators from the matter-ghost system. Therefore, we define our total BRST charge as
\begin{equation}
\begin{split}
Q_{\text{B}} &= Q_{\text{S}} + Q_{\text{V}}, \\
Q_{\text{S}} &= \oint \frac{d\widetilde{z}}{2\pi i} \left(\widetilde{G}^-_M - \widetilde{Q}_G \right) + \oint \frac{dz}{2\pi i} \left(G^+_M - Q_G \right), \\
Q_{\text{V}} &= \oint \frac{d\widetilde{z}}{2\pi i} \left( \widetilde{c} \widetilde{T}_M + \widetilde{\gamma} \widetilde{G}^+_M + \frac{1}{2} \left( \widetilde{c} \widetilde{T}_G + \widetilde{\gamma} \widetilde{G}_G \right) \right) + \oint \frac{dz}{2\pi i} \left( c T_M + \gamma G^-_M + \frac{1}{2} \left( c T_G + \gamma G_G \right) \right).
\end{split}
\end{equation}
Then we define the space of physical states in our theory as \cite{dislter-nelson-2, dvv}
\begin{equation}
\mathcal{H} := \{\ket{\mathcal{O}} \in \frac{\text{ker} Q_{\text{B}}}{\text{im} Q_{\text{B}}} \  |  \ (L_0 - \widetilde{L}_0)  \ket{\mathcal{O}} = (b_0 - \widetilde{b}_0)  \ket{\mathcal{O}} = 0 \},
\end{equation}
i.e. the elements of the cohomology of $Q_{\text{B}}$ annihilated by $L_0 - \widetilde{L}_0$ and $b_0 - \widetilde{b}_0$, where the subscript 0 denotes zero modes. The constraints aside from taking $Q_{\text{B}}$ cohomology, namely
\begin{equation}
(L_0 - \widetilde{L}_0)  \ket{\mathcal{O}} = (b_0 - \widetilde{b}_0)  \ket{\mathcal{O}} = 0,
\end{equation}
are usually called ``weak physical state conditions''.

So what are the observables here? Let's first consider the matter sector, namely the operators from topological sigma model. From \cite{witten-tft}, we know that in a pure topological sigma model (with no coupling to gravity), the physical observables are the chiral primary fields $\omega_i$ (inserted at the $i$-th puncture), which are in one-to-one correspondence with the elements in the De Rham cohomology of the target space. Let $\Omega_i$ be a chiral superfield with $\omega_i$ being its lowest component. Therefore, by our definition above we have physical states \cite{dislter-nelson-2, dvv}
\begin{equation}
\ket{\mathcal{O}_{n_{i}}} = \Omega_i \Gamma^{n_i} C \widetilde{C} \delta(D_+C) \delta(\widetilde{D}_-\widetilde{C}) \ket{0},
\end{equation}
where 
\begin{equation}
\Gamma := D_+ D_- D_+ C,
\end{equation}
and $\ket{0}$ is the Fock vacuum. The corresponding operators are 
\begin{equation}
\mathcal{O}_{n_{i}} = \Omega_i \Gamma^{n_i} C \widetilde{C} \delta(D_+C) \delta(\widetilde{D}_-\widetilde{C}),
\end{equation}
as usually from the operator-state correspondence. It is clear that these operators are in the so called $-1$ picture.

Let us now ask a question: what kind of quantity is the factor $\Gamma = D_+ D_- D_+ C$ that appeared in physical observables? Recall that the superghost field $C$ is a section of $\mathit{\Pi}\mathcal{D}_-$ (because $\mathcal{D}_+$ is trivial). Therefore, abstractly we see that $\Gamma$ is a section of a trivial bundle $\hat{\mathcal{O}}$ over worldsheet $\Sigma_T$.

In terms of local semirigid coordinates $(\widetilde{z}, z | \widetilde{\theta}^{\mp}, \theta^{\mp})$ on worldsheet $\Sigma_T$, we have
\begin{equation}
\begin{split}
\Gamma &= 2\partial_{z} \gamma + 2\theta^-\theta^+ \partial_{z} \partial_z \gamma, \\
&= 2\partial_{z} (\gamma + \theta^-\theta^+ \partial_z \gamma).
\end{split}
\end{equation}
Recall that the superghost field C has decomposition
\begin{equation}
C = c^z + \theta^- a + \theta^+ \gamma + \theta^- \theta^+ \partial_z c^z.
\end{equation} 
As usual, $c$ is a section of $\mathit{\Pi}T{\Sigma_T}_{\text{red}}$, where $T{\Sigma_T}_{\text{red}}$ the holomorphic tangent bundle of a Riemann surface. Note that $\theta^+$ is now a global (constant) scalar in our semirigid setup. Therefore, we see that $\gamma$ is now a section of $T{\Sigma_T}_{\text{red}}$. As such, the lowest component of $\Gamma$ is a section of $T{\Sigma_T}_{\text{red}} \otimes T^*{\Sigma_T}_{\text{red}}$, where $T^*{\Sigma_T}_{\text{red}}$ is the holomorphic cotangent bundle of ${\Sigma_T}_{\text{red}}$.

However, that is not quite the case. Note that a physical observable is only inserted at a particular point (a puncture) on $\Sigma$. As such, $\Gamma$ is only inserted (or defined) at those punctures. Let $p_i \in \Sigma_T$ be the $i$-th puncture on $\Sigma_T$. Then more precisely speaking $\Gamma$ is an element in the vector space $\hat{\mathcal{O}}_{p_i}$. This data defines a line bundle $\mathfrak{L}_i$ over the entire supermoduli space $\mathfrak{M}_{g,N}$, whose fiber at the point $[\Sigma_T]$ is simply $\hat{\mathcal{O}}_{p_i}$. Over the interior points $\mathfrak{M}_{g,N}$, corresponding to smooth curves, this bundle is clearly trivial. However, things are more complicated on the boundary of $\mathfrak{M}_{g,N}$.

We would like to distinguish between several ghost number symmetries in our theory. There are the usual $BC$ ghost number symmetries:
\begin{equation}
U(1)_{bc}, \ U(1)_{\beta\gamma}, \ U(1)_{\widetilde{b}\widetilde{c}}, \ U(1)_{\widetilde{\beta}\widetilde{\gamma}}.
\end{equation}
In addition, from topological sigma model we have a ghost number symmetry $U(1)_m$ from twisted fermions. It is clear that our physical observables have the following ghost charges under $(U(1)_m, \ U(1)_{bc}, \ U(1)_{\beta\gamma}, \ U(1)_{\widetilde{b}\widetilde{c}}, \ U(1)_{\widetilde{\beta}\widetilde{\gamma}})$:
\begin{equation}
(q_i, 1, n_i - 1, 1, -1).
\end{equation}
Now we can obtain a consequence from the expression of ghost supercurrent, which is inhomogeneous with respect to $U(1)_{bc}$ and $U(1)_{\beta\gamma}$. More precisely, we can only ask for the linear combination
\begin{equation}
U(1)_{bc} + 2U(1)_{\beta\gamma}
\end{equation}
to be conserved.

\subsection{Amplitudes}

We will define our topological string amplitudes as integration over our supermoduli space. We would like to proceed by ``picking a slice'', in other words choosing a set of local coordinates. Let $(\widetilde{\mathbf{m}}_{\tilde{k}}, \mathbf{m}_k) = (\widetilde{m}_{\tilde{k}}, m_k | \widetilde{\zeta}_{\tilde{k}}, \zeta_{k})$ be a set of local coordinates on our supermoduli space $\mathfrak{M}_{g,N}$. Let
\begin{equation}
\mathbf{V} = (\widetilde{V}_{\tilde{k}}, V_k | \widetilde{\Upsilon}_{\tilde{k}}, \Upsilon_k),
\end{equation} 
be a vector field on $\mathfrak{M}_{g,N}$, with even and odd components $(\widetilde{V}_{\tilde{k}}, V_k)$ and $(\widetilde{\Upsilon}_{\tilde{k}}, \Upsilon_k)$, respectively, using this local coordinate system. By definition, $\mathbf{V}$ represents the first order deformations of our topological string worldsheet $\Sigma$ represented by local coordinates $(\widetilde{\mathbf{m}}_{\tilde{k}}, \mathbf{m}_k)$. As discussed in appendix \ref{deformation}, $(\widetilde{V}_{\tilde{k}}, V_k)$ and $(\widetilde{\Upsilon}_{\tilde{k}}, \Upsilon_k)$ are elements of $H^1(\mathbf{S}, \mathcal{V}) \oplus H^1(\mathbf{S}, \widetilde{\mathcal{V}})$, i.e. they are one forms on $\Sigma$ valued in semirigid vector fields on $\Sigma$. 

To define topological string amplitudes, let's first define the field theory correlation function on a fixed semrigid super Riemann surface (represented by the point $(\widetilde{\mathbf{m}}_{\tilde{k}}, \mathbf{m}_k)$) as
\begin{equation}  \label{field-corr}
\mathcal{F}_{\mathcal{O}_{n_{1}} \mathcal{O}_{n_{2}} ... \mathcal{O}_{n_{N}}} := \int \mathcal{D} [\widetilde{\Phi}, \Phi, \widetilde{B}, \widetilde{C}, B, C] \  e^{-I} \ \prod_{i=1}^{N} \mathcal{O}_{n_{i}} \prod_{k=1}^{K} B_k \prod_{\tilde{k} = 1}^{K} \widetilde{B}_{\tilde{k}} \prod_{k=1}^{K} \delta(B_k) \prod_{\tilde{k}=1}^{K} \delta(\widetilde{B}_{\tilde{k}}),
\end{equation}
where again $\mathcal{D} [\widetilde{\Phi}, \Phi, \widetilde{B}, \widetilde{C}, B, C]$ denotes the path integral measure over all the fields in the theory, and we have superghost insertions:
\begin{equation}
\begin{split}
B_k & := \oint dz d\theta^- d\theta^+ B V_k, \\
\widetilde{B}_{\tilde{k}} & := \oint d\widetilde{z} d\widetilde{\theta}^- d\widetilde{\theta}^+ \widetilde{B} \widetilde{V}_{\tilde{k}}, \\
\delta(B_k) & := \delta\left( \oint dz_m d\theta^- d\theta^+ B \Upsilon_k \right), \\
\delta(\widetilde{B}_{\tilde{k}}) & := \delta\left(\oint d\widetilde{z} d\widetilde{\theta}^- d\widetilde{\theta}^+ \widetilde{B} \widetilde{\Upsilon}_{\tilde{k}} \right).
\end{split}
\end{equation} 
The contours in these formulas are those circling superghosts insertions counterclockwise.

What kind of quantity is $\mathcal{F}_{\mathcal{O}_{n_{1}} \mathcal{O}_{n_{2}} ... \mathcal{O}_{n_{N}}}$ in terms of geometry? To answer that question, let's first define an infinite dimensional supermanifold $\mathfrak{P}_{g,N}$, which is the space of semirigid super Riemann surfaces with genus $g$, $N$ marked points, and local semirigid coordinates at those punctures. By definition, $\mathcal{F}_{\mathcal{O}_{n_{1}} \mathcal{O}_{n_{2}} ... \mathcal{O}_{n_{N}}}$ is a function on $\mathfrak{P}_{g,N}$. Note $\mathfrak{P}_{g,N}$ has a natural fibration over our supermoduli space $\mathfrak{M}_{g,N}$. As usually described in the operator formalism of superstring theory (see, for example, \cite{operator-formalism}), $\mathcal{F}_{\mathcal{O}_{n_{1}} \mathcal{O}_{n_{2}} ... \mathcal{O}_{n_{N}}}$ can be thought of as the pull-back of a function on $\mathfrak{M}_{g,N}$, by picking a section $\sigma: \mathfrak{M}_{g,N} \rightarrow \mathfrak{P}_{g,N}$.

There are various constraint we have to impose on this data. First of all, standard results from pure topological gravity requires \cite{witten-ts}
\begin{equation} \label{selection}
\sum_{i=1}^N n_i= 3g - 3 + N.
\end{equation}
This condition guarantees the cancellation of ghost number anomaly from the linear combination $U(1)_{bc} + 2U(1)_{\beta\gamma}$ we discussed earlier. In addition, to cancel the anomaly from the ghost number anomaly from the matter sector, we must impose the following condition \cite{witten-tft}
\begin{equation} \label{selection2}
\sum_{i=1}^{N} q_i = d (1 - g),
\end{equation}
where $d$ is the complex dimension of the Calabi-Yau target space $M$.

Therefore, at this point it is natural to define our topological string amplitudes as 
\begin{equation} \label{amplitude}
\braket{\mathcal{O}_{n_{1}} \mathcal{O}_{n_{2}} ... \mathcal{O}_{n_{N}}}_g = \int_{\mathfrak{M}_{g,N}} \mathcal{D}(\widetilde{\mathbf{m}}, \mathbf{m}) \ \mathcal{F}_{\mathcal{O}_{n_{1}} \mathcal{O}_{n_{2}} ... \mathcal{O}_{n_{N}}},
\end{equation}
where
\begin{equation}
\mathcal{D}(\widetilde{\mathbf{m}}, \mathbf{m}) := [d\widetilde{m}_1...d\widetilde{m}_K, dm_1...dm_K | d\widetilde{\zeta}_1...d\widetilde{\zeta}_K, d\zeta_1...d\zeta_K]
\end{equation}
is an section of the Berezinian line bundle of $\mathfrak{M}_{g,N}$. 

Things are actually even simpler than that. As shown in appendix \ref{berezinian}, the Berezinian line bundle of $\mathfrak{M}_{g,N}$ is trivial. Therefore, there exists a global integration measure for the coordinates $(\widetilde{\mathbf{m}}_{\tilde{k}}, \mathbf{m}_k)$ on $\mathfrak{M}_{g,N}$, and this measure doesn't change across coordinate patches. In other words, there exists a globally well defined $\mathcal{D}(\widetilde{\mathbf{m}}, \mathbf{m})$. As such, $\mathcal{F}_{\mathcal{O}_{n_{1}} \mathcal{O}_{n_{2}} ... \mathcal{O}_{n_{N}}}$, which by definition is simply a function on $\mathfrak{M}_{g,N}$, can be naturally integrated over $\mathfrak{M}_{g,N}$ using such a global measure. Clearly our result is well defined and independent of the slice we have chosen.

Unlike the physical RNS string theory case \cite{witten-srs, witten-spt}, where one needs to construct a middle-dimensional cycle to properly define string amplitudes, this naive definition is the right way to proceed. The crucial point is that, just like when we define topological string worldsheets, we don't have any spinors in our topological theory. As such, we can indeed put reality conditions on anti-commuting moduli so our naive definition above is the correct one.

It is natural at this point to ask: what kind of quantity do these topological string amplitudes compute? As a priori, we would like to stress that these amplitudes are defined as integrations over supermoduli spaces (which are not projected in general). Therefore, the precise meaning of these amplitudes in terms of geometry is somewhat mysterious. In those very special cases where the supermoduli spaces involved are projected, we can consistantly integrate over odd moduli and the usual intersection theory interpretation of topological string theory amplitudes comes in. However, in general it is more complicated. The detailed discussion along these lines is left for future work.

\subsection{Picture Changing}

In supergeometry, as was argued in \cite{witten-sm, witten-spt}, it is in general not a good idea to try to reduce everything to the reduced space by integrating over odd coordinates. This originates from the fact that naturally there are mixings between even and odd variables under coordinate transformations. In superstring perturbation theory, naively integrating out odd moduli to obtain a measure on the bosonic moduli space often leads to difficulties such as spurious singularities. The best practice is to use what is naturally defined on supermanifolds and perform integrations over supermanifolds\footnote{In special cases, it might be viable to reduce to integrals over bosonic moduli spaces without encountering any trouble. See, for example, \cite{ohmori-tachikawa}.}.

With that said, however, to make connection with the traditional literature of topological string theory, we need to find a way to integrate out the odd modulus and reduce our topological string amplitudes to an integration over the bosonic moduli space. To do so, we need to find operators in a different picture number, which can be achieved by using the so-called picture changing operator:
\begin{equation}
P := Q_{\text{B}} \cdot \Theta(\beta) =  \left( G^-_M - b \right)\delta(\beta) - c \delta(\beta)\partial\beta,
\end{equation}
where $\Theta(\beta)$ is the step function. 

As in the case of superstring, picture changing operators are results from integrating out odd moduli on $\mathfrak{M}_{g,N}$ in the appropriate sense when it is possible to do so. In terms of supergeometry, we can obtain a basis for the odd moduli by using modes of gravitinos. By implementing gravitino modes with delta function support, we can obtain the above formula of picture changing operators (see \cite{witten-spt} for some more detailed discussion). The number of picture changing operators produced this way is precisely the number of odd moduli we have, namely $3g-3+N$.

Now let's try to integrate out the odd moduli using our picture changing operator $P$. Abstractly, we can simply rewrite the amplitude (\ref{amplitude}) as an expression involving only an integration over $\mathcal{M}_{g,N}$ with other factors suitably rewritten, by choosing a basis for the odd moduli and integrate them out, similar to the superstring case studied in detail in \cite{witten-spt}. Here in this paper we will pursue an indirect approach to work out this in series of indirect stages.

Therefore, in order to obtain a sensible topological string amplitude in terms of integration over the bosonic moduli space $\mathcal{M}_{g,N}$, we need insertions of operators in the -1 picture, the appropriate number of picture changing operators. The natural picture number is then $n_i - 1$. So we define
\begin{equation} 
O_{n_i} := \omega_i \gamma_0^{n_i} c \widetilde{c} \delta(\gamma) \delta(\widetilde{\gamma}) P^{n_i} \widetilde{P}^{n_i},
\end{equation}
inserted at the $n_i$-th puncture, where $\widetilde{P}$ is the complex conjugate of $P$. Clearly $O_{n_i}$ has picture number $n_i - 1$, and it is obtained from operators $\mathcal{O}_{n_i}$ by integrating out odd moduli and anticommuting coordinates on the worldsheet.

Now we can write down a worldsheet field theory correlation function (to replace what we had in (\ref{field-corr})) as
\begin{equation} 
F_{O_{n_{1}} O_{n_{2}} ... O_{n_{N}}} := \int \mathcal{D} (\widetilde{\phi}, \phi, \widetilde{\psi}, \psi, \widetilde{b}, b, \widetilde{c}, c, \widetilde{\beta}, \beta, \widetilde{\gamma}, \gamma) \  e^{-I} \ \prod_{i=1}^N O_{n_{i}} \prod_{k=1}^{K} b_k \prod_{\tilde{k}=1}^K \widetilde{b}_{\tilde{k}},
\end{equation}
where we now have ordinary ghost insertions
\begin{equation}
\begin{split}
b_k & := \oint dz \ b v_k, \\
\widetilde{b}_{\tilde{k}} & := \oint d\widetilde{z} \ \widetilde{b} \widetilde{v}_{\tilde{k}}.
\end{split}
\end{equation}
Here $\widetilde{v}_{\tilde{k}}$ and $v_k$ are antiholomorphic and holomorphic vector fields on the bosonic moduli space $\mathcal{M}_{g,N}$. Observe that the constraint (\ref{selection}) from the last section
\begin{equation}
\sum_{i=1}^N n_i= 3g - 3 +N
\end{equation}
assures us that we have just the right number of picture changing operators inserted.

Note that $F_{O_{n_{1}} O_{n_{2}} ... O_{n_{N}}}$ defined in this way is naturally a density on $\mathcal{M}_{g,N}$, a consequence from the expression of our picture changing operator $P$ which provides appropriate ghost insertions. As such, we can then define topological string amplitude as an integral over the bosonic moduli space $\mathcal{M}_{g,N}$:
\begin{equation} 
\braket{\mathcal{O}_{n_{1}} \mathcal{O}_{n_{2}} ... \mathcal{O}_{n_{N}}}_g = \int_{\mathcal{M}_{g,N}} \mathcal{D}(\tilde{m}, m) F_{O_{n_{1}} O_{n_{2}} ... O_{n_{N}}},
\end{equation}
where
\begin{equation}
\mathcal{D}(\tilde{m}, m) := [d\widetilde{m}_1...d\widetilde{m}_K;dm_1...dm_K]
\end{equation}
is the integration measure on $\mathcal{M}_{g,N}$. Therefore, our amplitudes is well defined as the integrand is now a top form on $\mathcal{M}_{g,N}$.

One can show that this result is precisely what we would get if we started with (\ref{amplitude}) and performed integration over odd moduli by choosing a specific basis. Whenever it is valid to consistently integrate over the odd moduli, e.g. when the supermoduli space $\mathfrak{M}_{g,N}$ is projected, this formula should be equivalent to the more general definition (\ref{amplitude}) defined on $\mathfrak{M}_{g,N}$. However, the basic definition (\ref{amplitude}) is clearly more general and less ad hoc.

\section*{Acknowledgement}
The author is greatly in debt to J. Distler, who has provided many guidance and help, as well as collaboration on follow-up papers. The author would also like to thank E. Sharpe and E. Witten for helpful discussions. This work is supported by the National Science Foundation grant PHY-1316033.

\appendix

\section{A Brief Review of Supermanifolds} \label{sm}

A supermanifold, like an ordinary manifold, can be defined by glueing together certain appropriate local data. Roughly speaking, a supermanifold $\mathbf{M}$ of dimension $n|m$ is a topological space that is locally isomorphic to the affine superspace $\mathbb{A}^{n|m}$ over a given field, with even dimension $n$ and odd dimension $m$. This way we can construct real supermanifolds (locally isomorphic to $\mathbb{R}^{n|m}$), complex supermanifolds (locally isomorphic to $\mathbb{C}^{n|m}$), and so on. For some comprehensive discussion of supermanifolds, see \cite{witten-sm, witten-srs, donagi-witten-proj}.

The entire space $\mathbf{M}$ can be constructed by patching together open subsets $U_{\alpha}$ that are isomorphic to open subsets of $\mathbb{A}^{n|m}$. One can parameterize each $U_{\alpha}$ by some local commuting coordinates $x^1, x^2, ... , x^n$ and anticommuting coordinates $\theta^1, \theta^2, ... , \theta^m$, then provide appropriate glueing data on the overlaps. More explicitly, on the overlap $U_{\alpha} \cap U_{\beta}$, we have even transition functions $f_{\alpha\beta}$ and odd transition functions $g_{\alpha\beta}$:
\begin{equation}
\begin{split}
x^i_{\alpha} & = f^i_{\alpha\beta} (x^1_{\beta}, ..., x^n_{\beta} | \theta^1_{\beta}, ..., \theta^m_{\beta}), \\ 
\theta^s_{\alpha} & = g^s_{\alpha\beta} (x^1_{\beta}, ..., x^n_{\beta} | \theta^1_{\beta}, ..., \theta^m_{\beta}),
\end{split}
\end{equation}
which satisfy the usual consistency conditions on double and triple overlaps. Notice that in general, both $f_{\alpha\beta}$ and $g_{\alpha\beta}$ are functions of all the even and odd coordinates.

If a supermanifold $\mathbf{M}$ can be patched together by transition functions $f_{\alpha\beta}$ and $g_{\alpha\beta}$, such that all the $f_{\alpha\beta}$'s do not dependent on any of those odd coordinates $\theta^1, \theta^2, ... , \theta^m$, then we say that $\mathbf{M}$ is projected. If $\mathbf{M}$ satisfy a further condition that all the $g_{\alpha\beta}$'s have only linear dependence on the odd coordinates $\theta^1, \theta^2, ... , \theta^m$, then we say $\mathbf{M}$ is split.

Somewhat more precisely, a supermanifold $\mathbf{M}$ is a locally ringed space, whose structure sheaf is a sheaf of $\mathbb{Z}/2$-graded supercommutative algebras over its reduced space. This is following the idea of defining ordinary manifolds using both the underlying spaces and the appropriate functions (continuous, smooth, analytic, ...) over them.

Let's first describe the local model. Let $V \to M$ be an odd vector bundle over an ordinary manifold $M$. Then a split supermanifold $\mathbf{M} (M,V)$ is the pair $(M,\mathcal{O}_{\mathbf{M}})$, where $\mathcal{O}_{\mathbf{M}}$ is the sheaf of $\mathcal{O}_M$ valued sections of $\wedge^{\bullet} V^{\vee}$, the exterior powers of the dual bundle of $V$. A general supermanifold is a supercommutative locally ringed space that is locally isomorphic to some $\mathbf{M} (M,V)$. $\mathbf{M}$ is called split if it is globally isomorphic to some $\mathbf{M} (M,V)$.

Let $J$ be the ideal of $\mathcal{O}_{\mathbf{M}}$ consisting of all nilpotents. The we can recover an ordinary manifold $M$ (which defines the base space of the local model $(M,V)$ of $\mathbf{M}$) with structure sheaf $\mathcal{O}_M = \mathcal{O}_{\mathbf{M}} / J$. We call $(M, \mathcal{O}_M)$ the reduced space of $\mathbf{M}$, denoted by $\mathbf{M}_{\text{red}}$. Clearly, $\mathbf{M}_{\text{red}}$ can be naturally embedded into $\mathbf{M}$, simply by setting all the odd coordinates to zero. If there is a projection from $\mathbf{M}$ to $\mathbf{M}_{\text{red}}$, then we call $\mathbf{M}$ projected.

Let's consider an example. The simplest supermanifold is the affine superspace $\mathbb{A}^{n|m}$ over some field. Here we can use a set of global coordinates $(x^1, x^2, ... , x^n | \theta^1, \theta^2, ... , \theta^m)$ to cover the entire $\mathbb{A}^{n|m}$. However, this is not the correct notion of a superspace that's usually used in physics.

In this note we will mainly consider complex supermanifolds, as well as cs manifolds, where cs stands for complex supersymmetric \cite{cs}. A cs manifold has a reduced space which is a real smooth manifold, but we are not allowed to ask for any reality conditions once we include the odd coordinates. This is because we will use Euclidean manifolds, on which we are not allowed to put any reality conditions on spinors. The local model of cs manifolds is usually denoted as $\mathbb{R}^{n*|m}$. This is what's used in physics as a superspace.

Let's consider another example: complex projective superspace $\mathbb{C}\mathbb{P}^{n|m}$. It is defined by a similar way of defining ordinary complex projective space $\mathbb{C}\mathbb{P}^n$, which is its reduced space. We start with the affine superspace $\mathbb{C}^{n+1|m}$, with local coordinates $(z^1, z^2, ... , z^{n+1} | \theta^1, \theta^2, ... , \theta^m)$. Then we define 
\begin{equation}
\mathbb{C}\mathbb{P}^{n|m} \equiv  \left( \mathbb{A}^{n+1|m} - \{ 0 \} \right) / \mathbb{C}^{*},
\end{equation}
where the $\mathbb{C}^{*}$ action is given by
\begin{equation}
(z^1, z^2, ... , z^{n+1} | \theta^1, \theta^2, ... , \theta^m) \to (\lambda z^1, \lambda z^2, ... , \lambda z^{n+1} | \lambda \theta_1, \lambda \theta_2, ... , \lambda \theta_m), \ \ \lambda \in \mathbb{C}^{*}.
\end{equation}
We can cover $\mathbb{C}\mathbb{P}^{n|m}$ by open subsets $U_i$, $i = 0, 1, ..., m$, on which $z^i \neq 0$. The local coordinates on $U_i$ are given by $(z^1/z^i, ..., z^{n+1}/z^i | \theta^1/z^i, ..., \theta^n/z^i)$, with the $i$-th term $z^i/z^i = 1$ removed.

A particular example is $\mathbb{C}\mathbb{P}^{1|2}$, whose reduced space is $\mathbb{C}\mathbb{P}^1$. We can cover $\mathbb{C}\mathbb{P}^{1|2}$ with two open subsets
\begin{equation}
\begin{split}
U(z^2 \neq 0): &\ z = \frac{z^1}{z^2}, \ \theta^1_U = \frac{\theta^1}{z^2}, \theta^2_U = \frac{\theta^2}{z^2}, \\
V(z^1 \neq 0): &\ w = \frac{z^2}{z^1}, \ \theta^1_V = \frac{\theta^1}{z^1}, \theta^2_V = \frac{\theta^2}{z^1},
\end{split}
\end{equation}
where $(z^1, z^2 | \theta^1, \theta^2)$ are the homogenous coordinates. Then it is clear that transition functions on $U \cap V$ are
\begin{equation}
\begin{split}
z &= \frac{1}{w}, \\
\theta^1_U &= \frac{\theta^1_V}{w},\\
\theta^2_U &= \frac{\theta^2_V}{w}.
\end{split}
\end{equation}
Therefore, we see explicitly that $\mathbb{C}\mathbb{P}^{1|2}$ is split. By a slight generalization, we see that every complex projective superspace is split.

\section{Berezinian Line Bundle} \label{berezinian}

Performing integration over supermanifolds is a bit trickier than over ordinary manifolds. On an ordinary manifold, what can be integrated are differential forms of top degree. However, the correspondingly defined differential forms on a supermanifold are not suitable for integration. If we parametrize a supermanifold $\mathbf{M}$ with local coordinates $(t^1,...,t^n|\theta^1,...,\theta^m)$, then the difficulty can be seen as the difficulty in integration of $d\theta^i$, which is an even variable. The way out is to define what's called integral forms \cite{witten-sm, witten-srs}, which are suitable for integration over supermanifolds. In a sense, one has to understand integration as an algebraic procedure.

An alternative but equivalent way of doing integration over supermanifolds is to use sections of Berezinian line bundles. Given a vector bundle $E$ on a supermanifold $M$, one can define what's called the Berezinian line bundle of $E$, denoted $\text{Ber}(E)$, as follows. Let $g_{\alpha\beta}$ be transition functions of $E$ on the overlap of two open subsets $U_{\alpha}$ and $U_{\beta}$ of $M$. Then $\text{Ber}(E)$ is defined by transition functions $\text{Ber}(g_{\alpha\beta})$, the Berezinian of the matrices $g_{\alpha\beta}$.

If $M$ is a real supermanifold, then a section of $\text{Ber}(T^*M)$, where $T^*M$ is the cotangent bundle of $M$, can be integrated over $M$. If $M$ is a complex supermanifold with dimension $n|m$, then a section of $Ber(T^*M)$, where $T^*M$ is the holomprphic cotangent bundle of $M$, can be integrated over a submanifold of dimension $\frac{n}{2}|m$. This definition of integration over supermanifolds is equivalent to the definition using integral forms, as argued in \cite{witten-sm, witten-srs}.

\subsection{Worldsheet}

It can be shown that there is a short exact sequence of vector bundles \cite{witten-srs} on a $\mathcal{N} = 2$ super Riemann surfaces $S$
\begin{equation}
0 \rightarrow \mathcal{D}_- \oplus \mathcal{D}_+ \rightarrow TS \rightarrow \mathcal{D}_- \otimes \mathcal{D}_+ \rightarrow 0.
\end{equation}
In other words, $D_-$, $D_+$ and $\{D_-, D_+\}$ generate $TS$. Note that the superconformal superfield $\mathcal{Y}(z|\theta^{\mp})$ we defined in (\ref{superconformal-vector}) can now be interpreted as sections of $TS / (\mathcal{D}_- \oplus \mathcal{D}_+) \cong \mathcal{D}_- \otimes \mathcal{D}_+$. This is relevant when we discuss superconformal ghosts.

The dual short exact sequence is 
\begin{equation}
0 \rightarrow (\mathcal{D}_- \otimes \mathcal{D}_+)^* \rightarrow T^*S \rightarrow (\mathcal{D}_- \oplus \mathcal{D}_+)^* \rightarrow 0.
\end{equation}
This follows from the fact that both $\mathcal{D}_-$ and $\mathcal{D}_-$ are integrable.

For any line bundle $L$, we have a canonical isomorphism $L^* \cong L^{-1}$. So the above sequence is equivalent to
\begin{equation}
0 \rightarrow (\mathcal{D}_- \otimes \mathcal{D}_+)^{-1} \rightarrow T^*S \rightarrow (\mathcal{D}_-^{-1} \oplus \mathcal{D}_+^{-1})\rightarrow 0,
\end{equation}
from which we can derive that
\begin{equation} 
Ber(T^*S) \cong Ber( (\mathcal{D}_- \otimes \mathcal{D}_+)^{-1} ) \otimes Ber(\mathcal{D}_-^{-1} \oplus \mathcal{D}_+^{-1}).
\end{equation}
Note that $(\mathcal{D}_- \otimes \mathcal{D}_+)^{-1}$ is an even line bundle, hence $Ber( (\mathcal{D}_- \otimes \mathcal{D}_+)^{-1} ) \cong (\mathcal{D}_- \otimes \mathcal{D}_+)^{-1}$. On the other hand
\begin{equation}
\begin{split}
Ber(\mathcal{D}_-^{-1} \oplus \mathcal{D}_+^{-1}) & \cong Ber(\mathcal{D}_-^{-1}) \otimes Ber(\mathcal{D}_+^{-1}), \\
& \cong \mathcal{D}_- \otimes \mathcal{D}_+,
\end{split}
\end{equation}
where we used the fact that $\mathcal{D}_-$ and $\mathcal{D}_+$ are odd line bundles. Therefore, we conclude that $Ber(T^*S)$ is canonically a trivial line bundle. In other words, any complex function on $S$ can be integrated over $S$. This is important when we write down Lagrangians.

\subsection{Supermoduli Space}

Let's now consider the Berezinian line bundle $\text{Ber}(\mathfrak{M}_{g})$ of our supermoduli space $\mathfrak{M}_{g}$ as a smooth supermanifold. We would like to prove that it is in fact a trivial line bundle. The first step is to use a trick we used in the main text of this paper, namely we think of $\mathfrak{M}_{g}$ as a middle-dimensional cycle diagonally embeded in two copies of it:
\begin{equation}
\mathfrak{M}_{g} \hookrightarrow \mathfrak{M}_g \times \mathfrak{M}_g.
\end{equation}
As we have argued in the main text, this means that, as a real supermanifold, $\mathfrak{M}_{g}$ has 
\begin{equation}
\text{Ber}(\mathfrak{M}_{g}) \cong Ber(\mathfrak{M}_{g}) \otimes Ber(\mathfrak{M}_{g}),
\end{equation}
where $Ber(\mathfrak{M}_{g})$ is the holomorphic Berezinian line bundle defined by the holomorphic cotangent bundle $T^*\mathfrak{M}_{g}$. Therefore, our goal now is to prove $Ber(\mathfrak{M}_{g})$ is a trivial line bundle.

Let $(m_k | \hat{m}_k)$ be a set of local coordinates on $\mathfrak{M}_{g}$. Then a canonical basis for $T^*\mathfrak{M}_{g}$ is given locally by $(dm_k | d\hat{m}_k)$. Note that the anticommuting directions don't have any topological effect as they are infinitesimal, we can simply consider the local model of $\mathfrak{M}_{g}$ whose reduced space is $\mathcal{M}_{g}$. We know from the main text that the local model of $\mathfrak{M}_{g}$ is given by $\mathit{\Pi}T\mathcal{M}_{g} \rightarrow \mathcal{M}_{g}$ on  $\mathcal{M}_{g}$. In other words, locally $\mathfrak{M}_{g}$ looks like the total space of the anticommuting tangent bundle of $\mathcal{M}_{g}$. In terms of local coordinates, this means that we can identify $\hat{m}_k$ with $dm_k$.

Now let $E$ and $F$ be vector bundles (over $\mathfrak{M}_{g}$) generated by basis $(\hat{m}_k)$ and $(d\hat{m}_k)$, respectively. Clearly, $E$ is an odd vector bundle of rank $0|3g-3$ and $F$ is an even vector bundle with rank $3g-3|0$. In our local mode, these are the odd and even part of $T^*\mathfrak{M}_{g}$:
\begin{equation}
T^*\mathfrak{M}_{g} = E \oplus F.
\end{equation}
Let $W$ be the transition matrix of $T^*\mathfrak{M}_{g}$. Let $A$ and $D$ be transition matrices of $E$ and $F$, respectively. Then evidently, in our local model, we have
\begin{equation}
\text{Ber}(W) = \frac{\text{det} A}{\text{det} D} = 1.
\end{equation}
Therefore, we conclude that $Ber(\mathfrak{M}_{g})$ is a trivial line bundle. 

This finishes our proof of the statement that the Berezinian line bundle $\text{Ber}(\mathfrak{M}_{g})$ of our supermoduli space $\mathfrak{M}_{g}$ as a smooth supermanifold is trivial. As such, any function can be integrated over $\mathfrak{M}_{g}$. This is how we defined topological string amplitudes in the main text.

\section{(0,2) Worldsheet}  \label{02}

Here, we will use Witten's technique to construct worldsheets with (0,2) superconformal symmetry. Let's consider
\begin{equation}
\Sigma_{(0,2)} \hookrightarrow S_L \times S_R,
\end{equation}
where now $S_L$ is again an ordinary Riemann surface, while $S_R$ is an $\mathcal{N}=2$ super Riemann surface discussed above. Now $\Sigma_{(0,2)}$ is a smooth submanifold of dimension $2|2$ which is close to the diagonal, representing the (0,2) worldsheet. 

Let's parametrize $S_L$ with local coordinate $\widetilde{z}$, and $S_R$ with local superconformal coordinates $(z|\theta^-,\theta^+)$. We define the ``antiholomorphic functions'' on $\Sigma_{(0,2)}$ to be holomorphic functions on $S_L$ restricted to $\Sigma_{(0,2)}$, and the ``holomorphic functions'' on $\Sigma_{(0,2)}$ to be holomorphic functions on $S_R$ restricted to $\Sigma_{(0,2)}$. In other words, we parametrize $\Sigma_{(0,2)}$ by $(\widetilde{z}, z| \theta^-,\theta^+)$ (restricted to $\Sigma_{(0,2)}$). Note we are not allowed to assume any reality condition on the odd coordinates here, as we are in Euclidean signature.

How do we define quantum field theories on $\Sigma_{(0,2)}$? Notice that a section of $Ber(S_L \times S_R) \cong Ber(S_L) \otimes Ber(S_R)$ \cite{witten-srs} can be integrated over a submanifold of $S_L \times S_R$ with dimension $1|2$, in particular over $\Sigma_{(0,2)}$. In fact, as a smooth supermanifold, the Berezinian line bundle (whose sections can be integrated over $\Sigma_{(0,2)}$) of $\Sigma_{(0,2)}$ is
\begin{equation}
\text{Ber}(\Sigma_{(0,2)}) \cong Ber(S_L \times S_R)|_{\Sigma_{(0,2)}}.
\end{equation}
Therefore, we should construct Lagrangians that are sections of $Ber(S_L) \otimes Ber(S_R)$. 

In appendix \ref{berezinian}, we showed that $Ber(S)$ is canonically trivial for any $\mathcal{N} = 2$ super Riemann surface $S$. So $Ber(S_R)$ is trivial. On the other hand, by definition $Ber(S_L)$ is simply the canonical line bundle $K_L \rightarrow S_L$, i.e. the bundle of top differential forms on $S_L$. We let $\Phi$ be a chiral function on $\Sigma_{(0,2)}$. Then
\begin{equation}
d\widetilde{z} \frac{\partial \Phi}{\partial \widetilde{z}} \in \Gamma(K_L),
\end{equation}
so we can construct the action for (0,2) chiral superfields as 
\begin{equation}
\int_{\Sigma_{(0,2)}} \mathcal{D}(\widetilde{z},z|\theta^-,\theta^+) \widetilde{\Phi}^i \partial_{\widetilde{z}} \Phi^i,
\end{equation}
where $\widetilde{\Phi}^i$ is a chiral function on $\Sigma_{(0,2)}$ that is the pullback from $X'$ (as in section \ref{srs}). This is of the same form as in the familiar case of (0,2) theories on flat $\mathbb{C}$. As mentioned before, since we are in Euclidean signature, we can't ask $\widetilde{\Phi}^i$ to be the complex conjugate of $\Phi^i$.

Next, we let $\Lambda$ be a section of $K^{1/2}_L$, i.e. we fix a spin structure on $S_L$. Then we call $\Lambda$ a (0,2) fermi superfield. The corresponding action is
\begin{equation}
\int_{\Sigma_{(0,2)}} \mathcal{D}(\widetilde{z},z|\theta^-,\theta^+) \widetilde{\Lambda}^a \Lambda^a,
\end{equation}
where $\widetilde{\Lambda}^a$ is also a section of $K^{1/2}_L$. Again, $\widetilde{\Lambda}^a$ is not the complex conjugate of $\Lambda^a$. They have expansion
\begin{equation}
\begin{split}
\Lambda & = \lambda + \theta^+ G - \theta^- \theta^+ \partial_z \lambda, \\
\widetilde{\Lambda} & = \widetilde{\lambda} + \theta^- \widetilde{G} + \theta^- \theta^+ \partial_z \widetilde{\lambda},
\end{split}
\end{equation}
where $\lambda$ and $\widetilde{\lambda}$ are left-moving Weyl fermionic fields, while $G$ and $\widetilde{G}$ are auxiliary fields.

One can further develop other types of (0,2) Lagrangians along these lines. For example, the gauge field strength superfield is simply a special fermi superfield whose lowest component is the left-moving gaugino. However, there is a caveat: we would like our theory to have the full (0,2) superconformal symmetry, since we with started with a worldsheet that enjoys this large symmetry. This puts further constraint on what kind of terms we can add the our Lagrangian, namely we would need the beta function of our theory to vanish. We will not discuss (0,2) theories further.

\section{$\mathcal{N}=2$ Deformation Theory} \label{deformation}

In this section, we discuss the details of how to characterize the deformations of a $\mathcal{N}=2$ worldsheet $\Sigma$. Recall that it is convenient to express $\Sigma$ as a middle dimensional cycle embedded as
\begin{equation}
\Sigma \hookrightarrow \mathbf{S}_L \times \mathbf{S}_R,
\end{equation}
This way, we can easily distinguish between antiholomorphic (left-moving) and holomorphic (right-moving) quantities. Because this is simply a bookkeeping trick, we see that everything on $\mathbf{S}_L$ (or $\mathbf{S}_R$), such as vector bundles $\mathcal{D}_{\mp}$ (or $\widetilde{\mathcal{D}}_{\mp}$), can be naturally restricted to $\Sigma$ without any changing of meanings.

Clearly, the complex vector bundle $T\mathbf{S}_L \oplus T\mathbf{S}_R$, when restricted to $\Sigma$, defines a what we call the tangent bundle $T\Sigma$, which is a complexification of the underlying real tangent bundle of $\Sigma$. We define 
\begin{equation}
T_L\Sigma := T\mathbf{S}_L|_{\Sigma}, \ \ \ \ T_R\Sigma := T\mathbf{S}_R|_{\Sigma}.
\end{equation}
Holomorphic sections of $T_L\Sigma$ are called antiholomorphic vector fields on $\Sigma$, while holomorphic sections of $T_R\Sigma$ are called holomorphic vector fields on $\Sigma$. In terms of local coordinates $(\widetilde{z}, z | \widetilde{\theta}^{\mp}, \theta^{\mp})$, we say a complex function $f$ on $\Sigma$ is holomorphic if 
\begin{equation}
\widetilde{D}_- f = \widetilde{D}_+ f = 0.
\end{equation}
Similarly, we say a complex function $g$ on $\Sigma$ is antiholomorphic if
\begin{equation}
D_- g = D_+ g = 0.
\end{equation}
The superghost fields we used in the main text are examples of this language.

We would like to study the first order deformations of the data defining $\Sigma$. There are essentially four ways to do that:
\begin{itemize}
\item Deforming the holomorphic structure of $\Sigma$, in other words deforming the embedding of $T_L\Sigma$ in $T\Sigma$.
\item Deforming the antiholomorphic structure of $\Sigma$, in other words deforming the embedding of $T_R\Sigma$ in $T\Sigma$.
\item Deforming the holomorphic superconformal structure of $\Sigma$, in other words deforming the embedding of $\mathcal{D}_{\mp}$ in $T_R\Sigma$.
\item Deforming the antiholomorphic superconformal structure of $\Sigma$, in other words deforming the embedding of $\widetilde{\mathcal{D}}_{\mp}$ in $T_L\Sigma$.
\end{itemize}

It can be shown, using techniques similar to what's used in \cite{witten-srs}, that the allowed deformations take the following form:
\begin{equation}
\begin{split}
\mathcal{D}_{\mp} &\rightarrow \mathcal{D}_{\mp} + \widetilde{h}_{\widetilde{z}}^z \partial_z + \widetilde{\chi}^-_{\widetilde{z}} \mathcal{D}_{-} + \widetilde{\chi}^+_{\widetilde{z}} \mathcal{D}_{+}, \\
\widetilde{\mathcal{D}}_{\mp} &\rightarrow \widetilde{\mathcal{D}}_{\mp} + h^{\widetilde{z}}_z \partial_{\widetilde{z}}  + \chi^-_{z} \widetilde{\mathcal{D}}_{-} + \chi^+_{z} \widetilde{\mathcal{D}}_{+},
\end{split}
\end{equation}
where $\widetilde{h}_{\widetilde{z}}^z$ and $h^{\widetilde{z}}_z$ are usually called metric perturbations, while $\widetilde{\chi}^{\mp}_{\widetilde{z}}$ and $ \chi^{\mp}_{z}$ are usually called gravitino fields. These worldsheet fields are defined up to gauge transformations:
\begin{equation}
\begin{split}
\widetilde{h}_{\widetilde{z}}^z &\rightarrow \widetilde{h}_{\widetilde{z}}^z + \partial_{\widetilde{z}} \widetilde{q}^{\widetilde{z}}, \ \ \ h^{\widetilde{z}}_z \rightarrow h^{\widetilde{z}}_z + \partial_z q^z, \\
\widetilde{\chi}^{\mp}_{\widetilde{z}} &\rightarrow \widetilde{\chi}^{\mp}_{\widetilde{z}} + \partial_{\widetilde{z}} \widetilde{\eta}^{\mp}, \ \ \chi^{\mp}_{z} \rightarrow \chi^{\mp}_{z} + \partial_z \eta^{\mp}.
\end{split}
\end{equation}
Therefore, these perturbations are cohomology classes on $\Sigma$ valued in semirigid vector fields. More precisely, they are elements of
\begin{equation}
H^1(\Sigma, \mathcal{V}) \oplus H^1(\Sigma, \widetilde{\mathcal{V}}),
\end{equation}
where $\mathcal{V}$ is the sheaf of holomorphic semirigid vector fields on $\Sigma$, and $\widetilde{\mathcal{V}}$ is the sheaf of antiholomorphic semirigid vector fields on $\Sigma$. Therefore, the holomorphic tangent space (or antiholomorphic tangent space) of the supermoduli space of semirigid super Riemann surfaces is isomorphic to $H^1(\Sigma, \mathcal{V})$ (or $H^1(\Sigma, \widetilde{\mathcal{V}})$).

We can reinterpret this result as deformations of complex structures on $\Sigma$. Let $\mathcal{J}$ be an endomorphism of $T\Sigma$ such that $\mathcal{J}^2 = -1$. Such a $\mathcal{J}$ is called an almost complex structure on $\Sigma$. In this language, $T_L\Sigma$ and $T_R\Sigma$ defined above are eigenbundles of $\mathcal{J}$ with eigenvalues $-i$ and $i$, respectively. $\mathcal{J}$ is called a complex structure if the sections of $T_L\Sigma$ and $T_R\Sigma$ separately form a Lie algebra. This is analogous to the usual integrability condition on ordinary manifolds. We would like to deform $\mathcal{J}$ in a fashion that it remains integrable.

The space of all such $\mathcal{J}$'s is an infinite-dimensional supermanifold. Let's call it $\mathfrak{P}$. Our worldsheet field theory path integral naturally defines a function on this large space. $\mathfrak{P}$ has a natural fibration over our supermoduli space $\mathfrak{M}$. Choosing a (local) section of this fibration is usually called ``picking a slice''. With such a choice, we call pull that function back to $\mathfrak{M}$ and define string amplitudes as an integral over $\mathfrak{M}$.

In general $\mathcal{N} = 2$ case, the integrability condition is always satisfied, a consequence of the $\mathcal{N} = 2$ superconformal structure on $\Sigma$. The nontrivial deformations are given by
\begin{equation}
\begin{split}
\delta \widetilde{\mathcal{J}}_{\widetilde{z}}^z &= \widetilde{h}_{\widetilde{z}}^z + \widetilde{\chi}^-_{\widetilde{z}} \theta_{-} + \widetilde{\chi}^+_{\widetilde{z}} \theta_{+}, \\
\delta \mathcal{J}^{\widetilde{z}}_z &= h^{\widetilde{z}}_z + \chi^-_{z} \widetilde{\theta}_{-} + \chi^+_{z} \widetilde{\theta}_{+},
\end{split}
\end{equation}
where $\widetilde{h}_{\widetilde{z}}^z$, $h^{\widetilde{z}}_z$, $\widetilde{\chi}^{\mp}_{\widetilde{z}}$ and $ \chi^{\mp}_{z}$ are as the above. On the worldsheet $\Sigma$, $\delta \widetilde{\mathcal{J}}_{\widetilde{z}}^z$ is a $(0,1)$ form with values in $\mathcal{D}_- \otimes \mathcal{D}_+$, while $\delta \mathcal{J}^{\widetilde{z}}_z$ is a $(1,0)$ form with values in $\widetilde{\mathcal{D}}_- \otimes \widetilde{\mathcal{D}}_+$.

Let $p$ be a point on $\Sigma$. Then geometrically it is natural to interpret $\delta \widetilde{\mathcal{J}}_{\widetilde{z}}^z(p)$ as a $(1,0)$ form on the infinite dimensional supermanifold $\mathfrak{P}$, and $\delta \mathcal{J}^{\widetilde{z}}_z(p)$ as a $(0,1)$ form on $\mathfrak{P}$. These forms on $\mathfrak{P}$ can be coupled to superghosts $\widetilde{B}$ and $B$ to provide approporiate superghost insertions needed in path integral as we saw in the main text.

\section{Amplitudes: An Alternative Approach} \label{alternative}

We will now define our topological string amplitudes using an alternative approach, based on integral forms. We will do this essentially because there are two ways to define integration on supermanifolds, and they are equivalent to each other. Let's first consider the deformations of complex structures $\delta \widetilde{\mathcal{J}}_{\widetilde{z}}^z, \delta \mathcal{J}^{\widetilde{z}}_z$ that preserve our semirigid structure.\footnote{Please see appendix \ref{deformation} for more details.} Note that $\delta \widetilde{\mathcal{J}}_{\widetilde{z}}^z$ and $\delta \mathcal{J}^{\widetilde{z}}_z$ can be thought of as $(1,0)$ and $(0,1)$ forms on our supermoduli space.

Let's consider the following coupling between $\delta \widetilde{\mathcal{J}}_{\widetilde{z}}^z$ and superghost $B$:
\begin{equation} \label{JB1}
\int_{X} \mathcal{D}(\widetilde{z},z|\theta^-,\theta^+)  \delta \widetilde{\mathcal{J}}_{\widetilde{z}}^z B_z,
\end{equation}
where $X$ is a submanifold of $\Sigma$ parametrized by local coordinates $(\widetilde{z},z|\theta^-,\theta^+)$ as we have used in defining superghost Lagrangian. We then expand $B$ as a sum of its zero modes and nonzero modes:
\begin{equation}
B = \sum_{k = 1, ..., K | 1, ..., K} \mathfrak{u}_k \mathfrak{B}_k + \sum_{l} \mathfrak{v}_{l} \mathfrak{B}'_l,
\end{equation}
where $\mathfrak{B}_k$ and $\mathfrak{B}'_l$ are zero modes and nonzero modes of $B$, respectively. Note that $\mathfrak{B}_k$ contains $K$ even parts and $K$ odd parts. Here $\mathfrak{u}_k$ and $\mathfrak{v}_{l}$ are just coefficients. Using this notation, the contribution from the zero modes of $B$ to the above coupling is
\begin{equation}
\sum_{k = 1, ..., K | 1, ..., K} \mathfrak{u}_k \Psi_{k},
\end{equation}
where we have defined
\begin{equation}
\Psi_k := \int_{X} \mathcal{D}(\widetilde{z},z|\theta^-,\theta^+)  \delta \widetilde{\mathcal{J}}_{\widetilde{z}}^z \mathfrak{B}_{zk}.
\end{equation}

A completely analogous story can be told for the antiholomorphic superghost $\widetilde{B}$. Now we define the following coupling:
\begin{equation} \label{JB2}
\int_{\widetilde{X}} \mathcal{D}(\widetilde{z},z|\widetilde{\theta}^-, \widetilde{\theta}^+) \delta \mathcal{J}^{\widetilde{z}}_z \widetilde{B}_{\widetilde{z}},
\end{equation}
where $\widetilde{X}$ is a submanifold of $\Sigma$ parametrized by local coordinates $(\widetilde{z},z|\widetilde{\theta}^-,\widetilde{\theta}^+)$ as we have used in defining superghost Lagrangian. Again, we can expand $\widetilde{B}$ using its zero modes and nonzero modes, and obtain the contribution form its zero modes $\widetilde{\mathfrak{B}}_k$
\begin{equation}
\sum_{k = 1, ..., K | 1, ..., K} \widetilde{\mathfrak{u}}_k \widetilde{\Psi}_{k},
\end{equation}
where we have defined
\begin{equation}
\widetilde{\Psi}_k := \int_{\widetilde{X}} \mathcal{D}(\widetilde{z},z|\widetilde{\theta}^-,\widetilde{\theta}^+)  \delta \mathcal{J}^{\widetilde{z}}_z \widetilde{\mathfrak{B}}_{\widetilde{z}k}.
\end{equation}

Now, on a give topolotical string worldsheet, we would like to compute the following worldsheet field theory correlation function
\begin{equation}  \label{field-corr1}
\begin{split}
F_{\mathcal{O}_{n_{1}} \mathcal{O}_{n_{2}} ... \mathcal{O}_{n_{N}}} := \int \mathcal{D} [\widetilde{\Phi}, \Phi, \widetilde{B}, \widetilde{C}, B, C] \ \prod_{i=1}^{N} \mathcal{O}_{n_{i}} \ \text{exp}\left(- I + \int_{X}  \delta \widetilde{\mathcal{J}}_{\widetilde{z}}^z B_z + \int_{\widetilde{X}} \delta \mathcal{J}^{\widetilde{z}}_z \widetilde{B}_{\widetilde{z}} \right),
\end{split}
\end{equation}
where $\mathcal{D} [\widetilde{\Phi}, \Phi, \widetilde{B}, \widetilde{C}, B, C]$ denotes the path integral measure over all the fields in the theory.

What kind of quantity is $F_{\mathcal{O}_{n_{1}} \mathcal{O}_{n_{2}} ... \mathcal{O}_{n_{N}}}$ on our supermoduli space $\mathfrak{M}_{g,N}$? The standard path integral over the ghost fields tells us the answer: the contribution from zero modes of $B$ is of the form
\begin{equation}
\prod_{k = 1, ..., K | 1, ..., K} \int d\mathfrak{u}_k \ \text{exp}(\mathfrak{u}_k \Psi_{k}) = \prod_{k = 1, ..., K | 1, ..., K} \delta(\Psi_{k}),
\end{equation}
while the contribution from the zero modes of $\widetilde{B}$ is of the form
\begin{equation}
\prod_{k = 1, ..., K | 1, ..., K} \int d\widetilde{\mathfrak{u}}_k \ \text{exp}(\widetilde{\mathfrak{u}}_k \widetilde{\Psi}_{k}) = \prod_{k = 1, ..., K | 1, ..., K} \delta(\widetilde{\Psi}_{k}).
\end{equation}
Therefore, by simply counting the number of superdegrees we see that $F_{\mathcal{O}_{n_{1}} \mathcal{O}_{n_{2}} ... \mathcal{O}_{n_{N}}}$ is an integral form on $\mathfrak{M}_{g,N}$ with superdegree $2K | 2K$. Therefore, it can be integrated over $\mathfrak{M}_{g,N}$ naturally, which is what we are going to do.

So let's define topological string amplitudes as
\begin{equation} \label{amplitude1}
\braket{\mathcal{O}_{n_{1}} \mathcal{O}_{n_{2}} ... \mathcal{O}_{n_{N}}}_g = \int_{\mathfrak{M}_{g,N}} F_{\mathcal{O}_{n_{1}} \mathcal{O}_{n_{2}} ... \mathcal{O}_{n_{N}}}.
\end{equation}
This definition is very natural, as $F_{\mathcal{O}_{n_{1}} \mathcal{O}_{n_{2}} ... \mathcal{O}_{n_{N}}}$ is naturally an integral form on $\mathfrak{M}_{g,N}$ with top degree. We don't need to rely on any local description of $\mathfrak{M}_{g,N}$ using local coordinates.

Unlike the physical RNS string theory case \cite{witten-srs, witten-spt}, where one needs to construct a middle-dimensional cycle to properly define string amplitudes, this naive definition is the right way to proceed. The crucial point is that, just like when we define topological string worldsheets, we don't have any spinors in our topological theory. As such, we can indeed put reality conditions on anti-commuting moduli so our naive definition above is the correct one.

Now we would like to claim that
\begin{equation}
\mathcal{D}(\widetilde{\mathbf{m}}, \mathbf{m}) \mathcal{F}_{\mathcal{O}_{n_{1}} \mathcal{O}_{n_{2}} ... \mathcal{O}_{n_{N}}} = \int \mathcal{D}(d\widetilde{\mathbf{m}}, d\mathbf{m}) F_{\mathcal{O}_{n_{1}} \mathcal{O}_{n_{2}} ... \mathcal{O}_{n_{N}}},
\end{equation}
where 
\begin{equation}
\mathcal{D}(\widetilde{\mathbf{m}}, \mathbf{m}) := [d\widetilde{m}_1...d\widetilde{m}_K;dm_1...dm_K | d\widetilde{\zeta}_1...d\widetilde{\zeta}_K;d\zeta_1...d\zeta_K],
\end{equation}
and $\mathcal{D}(d\widetilde{\mathbf{m}}, d\mathbf{m})$ are integral measures on coordinates and their differentials, respectively, using this particular slice we have chosen. That is to say, when we first perform the (algebraic) integration on $F_{\mathcal{O}_{n_{1}} \mathcal{O}_{n_{2}} ... \mathcal{O}_{n_{N}}}$ over the differentials $(d\widetilde{\mathbf{m}}_{\tilde{k}}, d\mathbf{m}_k)$, we obtain $\mathcal{D}(\widetilde{\mathbf{m}}, \mathbf{m})\mathcal{F}_{\mathcal{O}_{n_{1}} \mathcal{O}_{n_{2}} ... \mathcal{O}_{n_{N}}}$ as defined in the main text.

How do we see this? Let's recall that $\delta \widetilde{\mathcal{J}}_{\widetilde{z}}^z$ and $\delta \mathcal{J}^{\widetilde{z}}_z$ are $(1,0)$ and $(0,1)$ forms on our supermoduli space $\mathfrak{M}_{g,N}$. Using the particular slice we have chosen in terms of local coordinates $(\widetilde{\mathbf{m}}_{\tilde{k}}, \mathbf{m}_k)$, we can express $\delta \widetilde{\mathcal{J}}_{\widetilde{z}}^z$ and $\delta \mathcal{J}^{\widetilde{z}}_z$ as
\begin{equation}
\begin{split}
\delta \widetilde{\mathcal{J}}_{\widetilde{z}}^z &= \sum_{k=1, ..., K | 1, ..., K} \frac{\partial \widetilde{\mathcal{J}}_{\widetilde{z}}^z}{\partial \mathbf{m}_k} d \mathbf{m}_k, \\
\delta \mathcal{J}^{\widetilde{z}}_z &= \sum_{\tilde{k}=1, ..., K | 1, ..., K} \frac{\partial \mathcal{J}^{\widetilde{z}}_z}{\partial \widetilde{\mathbf{m}}_{\tilde{k}}} d \widetilde{\mathbf{m}}_{\tilde{k}}.
\end{split}
\end{equation}
Note we have utilized our explicit basis $(d\widetilde{\mathbf{m}}_{\tilde{k}}, d\mathbf{m}_k)$ of differentials.

Now lets rewrite the couplings we introduced in equations (\ref{JB1}) and (\ref{JB2}) using this basis:
\begin{equation}
\sum_{k=1, ..., K | 1, ..., K}  d\mathbf{m}_k \mathbf{B}_k + \sum_{\tilde{k}=1, ..., K | 1, ..., K}  d\widetilde{\mathbf{m}}_{\tilde{k}} \widetilde{\mathbf{B}}_{\tilde{k}},
\end{equation}
where we have defined
\begin{equation}
\begin{split}
\mathbf{B}_k &:= \int_{X} \mathcal{D}(\widetilde{z},z|\theta^-,\theta^+)  \frac{\partial \widetilde{\mathcal{J}}_{\widetilde{z}}^z}{\partial \mathbf{m}_k} B_z, \\
\widetilde{\mathbf{B}}_{\tilde{k}} &:= \int_{\widetilde{X}} \mathcal{D}(\widetilde{z},z|\widetilde{\theta}^-, \widetilde{\theta}^+) \frac{\partial \mathcal{J}^{\widetilde{z}}_z}{\partial \widetilde{\mathbf{m}}_{\tilde{k}}} \widetilde{B}_{\widetilde{z}}.
\end{split}
\end{equation}
Then the integration over the differentials $(d\widetilde{\mathbf{m}}_{\tilde{k}}, d\mathbf{m}_k)$ leads us to
\begin{equation}
\begin{split}
\prod_{\tilde{k}=1, ..., K | 1, ..., K} \int \mathcal{D}(d\widetilde{\mathbf{m}}_{\tilde{k}}) &\text{exp}\left( -d\widetilde{\mathbf{m}}_{\tilde{k}} \widetilde{\mathbf{B}}_{\tilde{k}} \right) \prod_{k=1, ..., K | 1, ..., K} \int \mathcal{D}(d\mathbf{m}_k) \text{exp} \left( - d\mathbf{m}_k \mathbf{B}_k \right), \\
&= \prod_{\tilde{k}=1, ..., K | 1, ..., K} \delta(\widetilde{\mathbf{B}}_{\tilde{k}}) \prod_{k=1, ..., K | 1, ..., K} \delta(\mathbf{B}_k).
\end{split}
\end{equation}

To relate what we just obtained in terms of vector fields $(\widetilde{V}_{\tilde{k}}, V_k | \widetilde{\Upsilon}_{\tilde{k}}, \Upsilon_k)$ on $\mathfrak{M}_{g,N}$, we simply apply super Beltrami equations
\begin{equation}
\mu_k = \partial_{\widetilde{z}}  (V_k | \Upsilon_k), \ \  \widetilde{\mu}_{\tilde{k}} = \partial_z (\widetilde{V}_{\tilde{k}} | \widetilde{\Upsilon}_{\tilde{k}}),
\end{equation}
where 
\begin{equation}
\mu_k := \frac{\partial \widetilde{\mathcal{J}}_{\widetilde{z}}^z}{\partial \mathbf{m}_k}, \ \ \widetilde{\mu}_{\tilde{k}} := \frac{\partial \mathcal{J}^{\widetilde{z}}_z}{\partial \widetilde{\mathbf{m}}_{\tilde{k}}}.
\end{equation}

Note that the $\delta$ function of an odd variable is simply that variable itself. Therefore, after performing integration by parts we see
\begin{equation}
\begin{split}
\prod_{k=1, ..., K | 1, ..., K} \delta(\mathbf{B}_k) & = \prod_{k=1}^{K} B_k \prod_{k=1}^{K} \delta(B_k), \\
\prod_{\tilde{k}=1, ..., K | 1, ..., K} \delta(\widetilde{\mathbf{B}}_{\tilde{k}}) &= \prod_{\tilde{k} = 1}^{K} \widetilde{B}_{\tilde{k}} \prod_{\tilde{k}=1}^{K} \delta(\widetilde{B}_{\tilde{k}}), 
\end{split}
\end{equation}
which justifies the superghost insertions we have defined. Therefore, we have now proven our claim.

\end{document}